\documentclass[amsmath,amssymb,pra,twocolumn]{revtex4-1}

\usepackage{graphics}
\usepackage{epsfig}
\usepackage{amsmath}
\usepackage{color}
\usepackage{dcolumn}
\usepackage{bm}
\usepackage{multirow}
\usepackage{amstext,amssymb,amsfonts,amsmath}
\usepackage{mathrsfs}
\usepackage{xspace}

\def\oper#1{\hat{\rm #1}}

\newcommand{\asc}{\ensuremath{a}}
\newcommand{\abg}{\ensuremath{a_{\rm bg}} }
\newcommand{\aho}{\ensuremath{a_{\rm ho}} }
\newcommand{\rint}{\ensuremath{r_{\rm int}} }
\newcommand{\cm}{\rm COM\ }
\newcommand{\rel}{\rm REL\ }
\newcommand{\cmi}{\rm COM\xspace}
\newcommand{\reli}{\rm REL\xspace}

\newcommand{\ket}[1]{\left| #1 \right>} 
\newcommand{\bra}[1]{\left< #1 \right|} 
\newcommand{\braket}[2]{\left< #1 \vphantom{#2} \right|
 \left. #2 \vphantom{#1} \right>} 
\newcommand{\QMa}[3]{\left< #1 \right| #2 \left| #3 \right>} 

\begin{document}

\bibliographystyle{/usr/share/texmf/tex/latex/revtex41/bibtex/bst/revtex/apsrev4-1}

  \title{Two-channel Bose-Hubbard model of atoms at a Feshbach resonance}

       \author{Philipp-Immanuel Schneider and Alejandro Saenz}

       \affiliation{AG Moderne Optik, Institut f\"ur Physik,
         Humboldt-Universit\"at zu Berlin, Newtonstra\ss e 15,
         12489 Berlin, Germany}

       \date{\today}

  \begin{abstract}
    Based on the analytic model of Feshbach resonances in harmonic traps described
    in Phys.\ Rev.\ A {\bf 83}, 030701 (2011)
    a Bose-Hubbard model is introduced that provides an accurate description of
    two atoms in an optical lattice at a Feshbach resonance with only a small number
    of Bloch bands. The approach circumvents the problem that the eigenenergies
    in the presence of a delta-like coupling do not converge to the correct energies, if an
    uncorrelated basis is used.
    The predictions of the Bose-Hubbard model are compared to non-perturbative calculations
    for both the stationary states and the time-dependent wavefunction during an acceleration
    of the lattice potential. 
    For this purpose, a square-well interaction potential is introduced, which allows for a
    realistic description of Feshbach resonances within non-perturbative single-channel
    calculations.
  \end{abstract}

    \maketitle

 \section{Introduction}
 \label{sec:intro}

Since the creation of the first Bose-Einstein
condensates~\cite{cold:ande95,cold:davi95}, ultracold atoms have proven to be a
versatile tool for many applications like precision measurement, quantum
simulation, and quantum information processing. Two of the main techniques that
made these achievements possible are the creation of various trapping
potentials, like optical lattices (OLs) or wave guides and the precise control of the
interatomic interaction by means of Feshbach resonances (FRs) \cite{cold:bloc08,cold:chin10}.

An important tool for describing ultracold atoms in OLs is the Bose-Hubbard (BH) model.
The model uses in its basic form a basis of single-particle Wannier states from the first Bloch
band to formulate the many-body Hamiltonian. 
While for weak interactions the model is very accurate, it usually breaks down for larger 
scattering lengths. A way to extend its applicability at a broad FR is to introduce effective
BH parameters especially for the onsite interaction strength $U$.
These parameters can be obtained by using a corrected harmonic approximation of the 
lattice sites \cite{cold:schn09} or by full numerical calculations \cite{cold:buch10,cold:buch12}.

The usual BH model allows via the onsite-interaction strength $U$ either for 
repulsively interacting atoms ($U>0$) or attractively interacting atoms ($U<0$).
At a narrow FR, however, a relatively narrow avoided crossing with the resonant bound
state leads to the appearance of both repulsively and attractively interacting states 
\cite{cold:schn11,cold:sand11}. In this situation the resonant bound state
must be explicitly included into the BH model. 
Several different kinds of these extended models have been introduced and debated 
\cite{cold:dick05,cold:dien06,cold:dick06,cold:sand11} and applied to map out the phase diagramm
\cite{cold:dick05,cold:carr05,cold:rous09} or to investigate lattice solitons \cite{cold:krut06}. 

The above investigations consider the extended Hubbard model within a single-band approximation
that is only applicable in the rare situation that the coupling energy to the resonant
bound state is small compared to the band gap. In order to generalize the applicability 
one can introduce the notion of dressed molecules with
effective bound-state energies and coupling strengths obtained from more elaborate calculations
\cite{cold:wall12}.

A convenient approach to generalize Hubbard models to describe broader FRs or
systems with a large scattering length is to simply include more Bloch bands. 
For example, L.-M. Duan has derived an effective single-band Hubbard model for the case of interacting 
fermions at a broad FR starting from a multi-band Hubbard model in the Wannier basis 
and a zero-range coupling between atoms and molecules \cite{cold:duan05}. 
However, as will be discussed in this work, severe numerical problems arise for the description
of a system with a zero-range coupling, e.g., by expanding the solution in products of
single-particle basis functions. 
Especially for large scattering lengths all of these basis functions behave completely differently for 
$r\rightarrow 0$ compared with the correct solution. 
This poses a problem especially for positive scattering lengths where the open channel supports a bound state.
In fact, the obtained energies are lower than the correct ones so 
that an increase of the basis leads to an even larger disagreement.
A similar problem also appears when replacing the interaction potential by the delta-like Fermi-Huang 
pseudo-potential~\cite{cold:esry99}.
Also within analytical treatments of FRs in harmonic traps that use non-interacting basis states
the eigenenergies do not converge~\cite{cold:dick05,cold:krzy13}. In this case, after an infinite summation, 
the diverging terms can be absorbed by introducing a renormalized bound-state energy.
In many numerical approaches the problem is circumvented by replacing the delta-like potential
by a regularized short range potential \cite{cold:buch10,cold:buch12,cold:brou12}. 
In order to resolve the potential usually a large basis is
necessary. For example, for an interaction with the range $d/N$ where $d$ is the lattice spacing
more than $N$ Bloch bands have to be included to converge the energies \cite{cold:buch10,cold:buch12}.
Since for two atoms in a one dimensional lattice the number of basis functions scales quadratically with 
the number of Bloch bands and the number of sites the solution can quickly become numerically 
very demanding.
Based on this corrected numerical approach, M.L. Wall and L.D. Carr were able to calculated the
effective parameters of a Fermi Hubbard model that takes the coupling to a bosonic molecule 
explicitly into account \cite{cold:wall12}.

In this work we introduce an extended BH model that avoids the numerical problems in the presence
of a delta-like coupling without the need of regularization and inclusion of many Bloch bands.
The model is derived from first principles on the basis of the analytic microscopic theory of FRs 
in a harmonic trap \cite{cold:schn11}. 
This allows for defining dressed bound-state energies and couplings that correct for the problems 
due to the deficiency of the basis states.

Given the number of different proposals to describe FRs within a BH model one has to compare
the predictions of the introduced BH model with non-perturbative calculations.
In the standard description of FRs this requires to solve a two-channel problem of two interacting 
atoms in an optical lattice coupled at short distance to a molecular bound state.
This problem is numerically very demanding. However, we show that one can largely simplify the
problem by introducing a square-well interaction potential that realistically mimics the
behavior at a FR. 
Using this single-channel interaction potential we apply an approach introduced in \cite{cold:gris11,cold:schn12a} 
in order to obtain the correct energies and wave functions of two atoms in a 
small OL at a FR.
The correct stationary and dynamic behavior of two atoms in a double-well potential is compared with 
the results of the introduced BH model.
It is shown that with only a small number of Bloch bands included the BH model is able to accurately 
describe FRs of small and medium width with coupling energies up to the depth of the OL.

The work is organized as follows. 
First the analytic model of a FR in a harmonic introduced in \cite{cold:schn11} is briefly recapitulated. 
This sets the basis for the derivation of a general BH model of interacting atoms at a FR in 
Sec.~\ref{sec:FRinOL}.  
The model is compared to the exact analytical solution in a harmonic trap in Sec.~\ref{sec:Problem}, 
revealing that the BH model does not converge toward the correct eigenenergies.
To circumvent this problem dressed molecular states and a dressed 
coupling strength of the BH model are introduced in Sec.~\ref{sec:dressing}
on the basis of the analytically known eigenenergies in the harmonic trap.
In Sec.~\ref{sec:numMethod} the square-well interaction potential is discussed, which allows for finding within
a non-perturbative approach both the stationary and the time-dependent wavefunctions of two atoms in a small OL at a FR 
perturbed by a time-dependent acceleration of the lattice.
Finally, in Sec.~\ref{sec:Comparison} the dressed and undressed BH model is compared to the non-perturbative 
calculations. We conclude in Sec.~\ref{sec:Conclusion}.

\section{Feshbach resonance in a harmonic trap}
\label{sec:FRinHarmTrap}

Neutral atoms usually only interact at small distances $r$ on the order of $\rint \sim 100\,$a.u.\ 
which is much smaller than typical length scales of the trapping potentials on the order of some 
$r_{\rm trap} \sim 10\,000\,$a.u. The collision energy in the ultracold regime is so small 
that partial waves with angular momentum $l>0$ are reflected by the centrifugal barrier.
Therefore $s$-wave ($l=0$) scattering is largely dominant. For $\rint\ll r\ll r_{\rm trap}$ the 
interaction leads to a phase shift $\varphi$ of the scattering wave function $\propto\sin(k r + \varphi)$ 
which is associated with the $s$-wave scattering length $\asc(k) = -\tan(\varphi)/k$.

\begin{figure}[ht]
   \centering
   \includegraphics[width=0.9\linewidth]{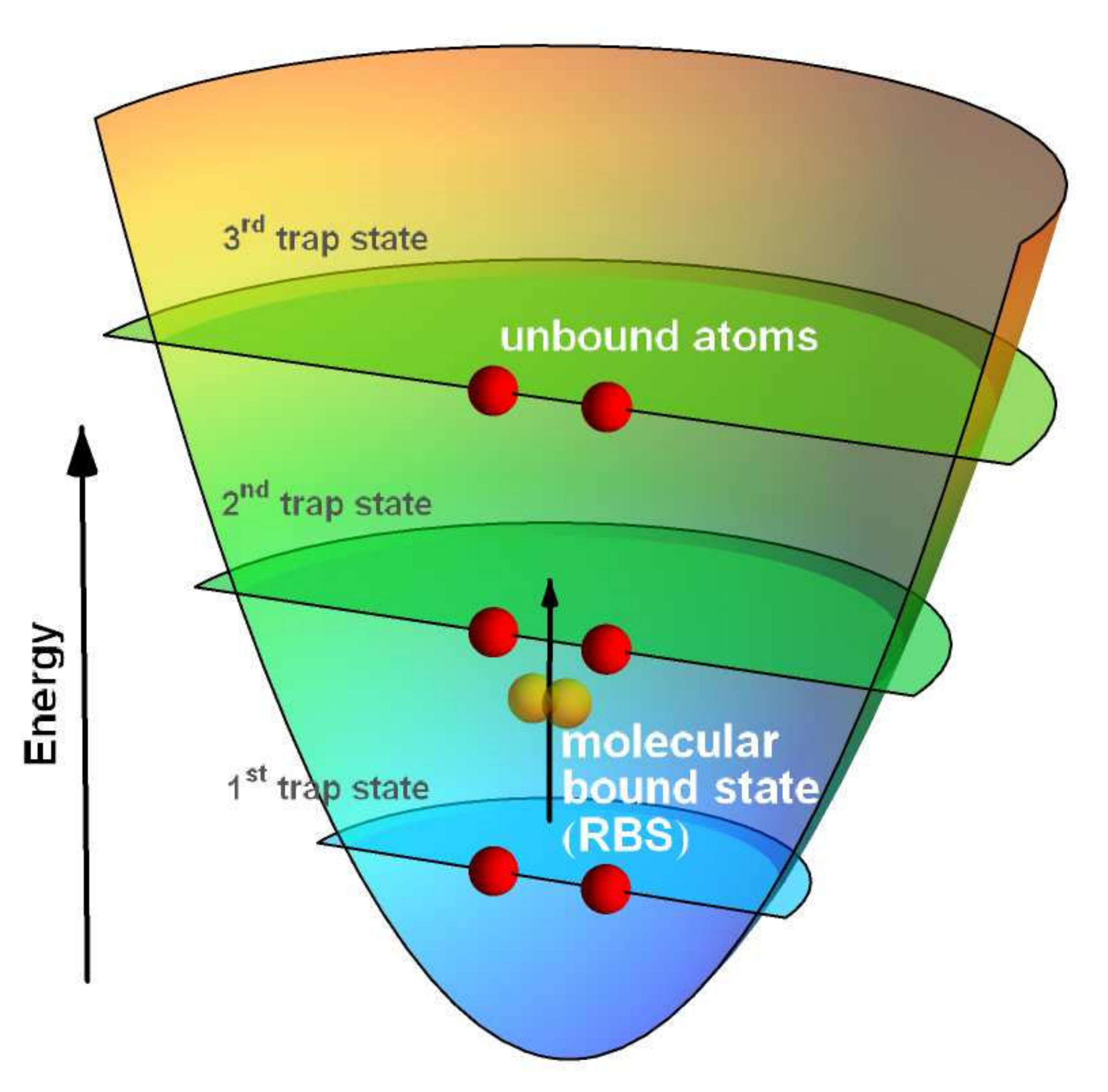}
   \includegraphics[width=0.98\linewidth]{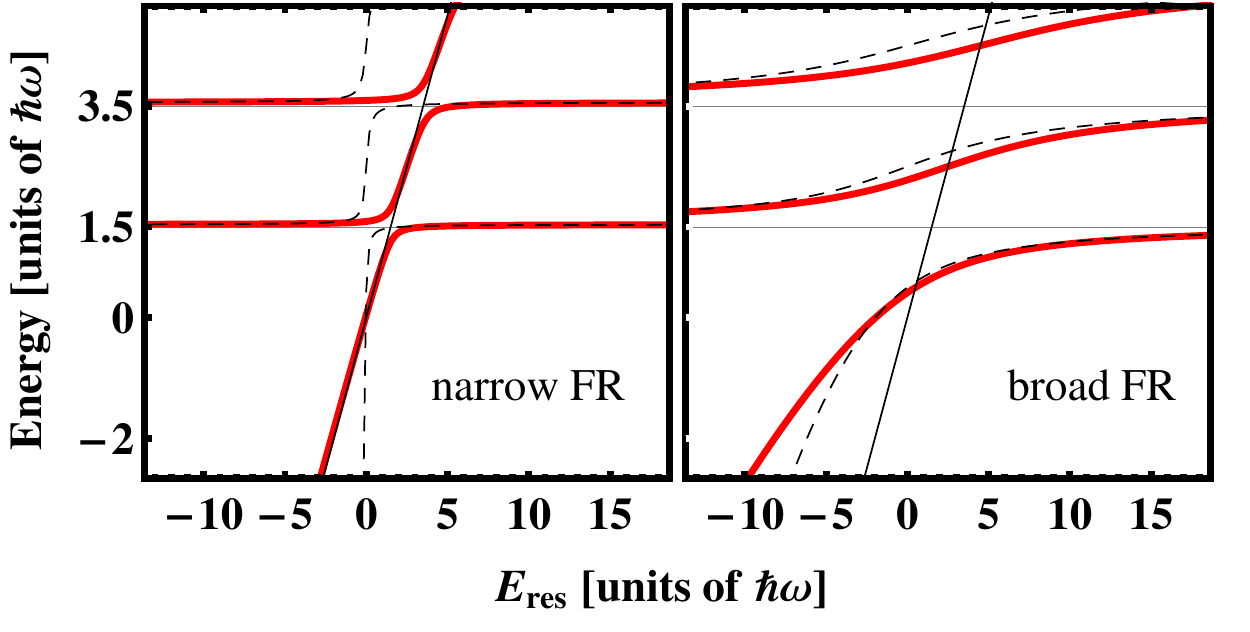}
   \caption{(Color online). Mechanism of broad and narrow FRs in the exemplary case of two atoms in  
   isotropic harmonic confinement with frequency $\omega$. 
   {\bf Top}: Sketch of the relative-motion unbound trap states and the resonant molecular bound state (RBS) whose
   energy can be manipulated by an external magnetic field.
   {\bf Bottom}: Relative-motion energy spectrum [solutions of Eq.~\eqref{eq:a_E_trap}] using the energy-dependent 
   scattering length $\asc(E,E_{\rm res})$ (thick red lines) and the energy-independent 
   scattering length $\asc(0,E_{\rm res})$ (black dashed line) as a function of the resonance 
   energy $E_{\rm res}=E_b + \delta E$ (black solid line). At a narrow FR (left: $\abg = 0.04\aho,
   \Delta E = 1\hbar\omega$) the RBS couples only to the trap state that is in resonance, which 
   leads to narrow avoided crossings. At a broad FR (right: $\abg = 0.04\aho, \Delta E = 40\hbar\omega$) 
   the RBS couples to many trap states and the energy spectrum changes globally with $E_{\rm res}$. 
   In contrast to the narrow FR the eigenenergies for an energy-dependent and energy-independent scattering 
   length agree reasonably for broad resonances.}
   \label{fig:SketchBroadVsNarrow}
  \end{figure}

If two atoms collide, the spin states of the scattering atoms are coupled at small 
distances $r\leq \rint$ to other spin states in closed channels whose relative energy can be influenced 
by applying an external magnetic field $B$. The subspace of closed-channel spin states can support many 
bound states. For certain magnetic field strengths $B$ the energy of such a bound state $E_b(B)$ can be 
brought into resonance with the collision energy $E$ of the atoms, leading to a FR
(see Fig.~\ref{fig:SketchBroadVsNarrow}).

In \cite{cold:schn11} an analytic model for a FR in isotropic and anisotropic harmonic traps was developed.
Its starting point is the relative-motion (\reli) Hamiltonian for radial momentum $l=0$ of two atoms in a
spherical harmonic confinement with frequency $\omega$.
The Hamiltonian for the radial wave function $\ket{\Phi(r)} = \sqrt{4\pi} r\ket{\Psi(r)}$, where $\ket{\Psi(r\,)}$
is the \rel wave function, is given as
\begin{equation}
\oper H = -\frac{\hbar^2}{2 \mu}\frac{d^2}{d r^2} + \frac12 \mu \omega^2 r^2 + \oper V_{\rm ZH}
         + \oper V_{\rm int}(r).
\end{equation}
 Here, $\mu$ is the reduced mass, $\oper V_{\rm ZH}$ is the Zeeman and hyperfine energy of the atoms, and 
 $\oper V_{\rm int}(r)$ the electron-spin dependent interaction potential.

 One assumes that the \rel energy of the atoms is small enough so that only one spin configuration
 (the open channel) supports unbound states. All other spin configurations are closed, i.e.\ their
 wave function vanishes for large interatomic distances.

 The analytic model is based on the two-channel description of an FR where one channel represents
 the unbound atoms and the other channel the subspace of closed channels.
 By introducing projectors $\oper P$ and $\oper Q$ onto the subspace of open and closed channels,
 respectively, one arrives at the coupled equations
 \begin{eqnarray}
 \label{eq:two_channel_1}
  (\oper H_P -E) \ket{\Phi_P} + \oper W \ket{\Phi_Q} &=& 0 \,,\\
 \label{eq:two_channel_2}
  (\oper H_Q -E) \ket{\Phi_Q} + \oper W^\dagger \ket{\Phi_P} &=& 0\,, 
 \end{eqnarray}
 with $\oper H_P = \oper P \oper H \oper P$, $\oper H_Q = \oper Q \oper H \oper Q$,
 $\oper W = \oper P \oper H \oper Q$, $\ket{\Phi_P}=\oper P \ket{\Phi}$,
 $\ket{\Phi_Q}=\oper Q \ket{\Phi}$, and $E$ the energy above the threshold of the
 open-channel interaction potential.
 Since the eigenenergies in the closed channel subspace are usually largely separated on the energy scale
 $\hbar \omega$ of the trap, one assumes that close to the FR $\ket{\Phi_Q}$ is simply a multiple $A$ 
 of a single bound eigenstate $\ket{\phi_b}$ with eigenenergy $E_b$. 
 We call this closed-channel state ``resonant bound state'' (RBS). To first order, the energy $E_b$ may be expanded 
 linearly in the magnetic field $B$, i.e. $E_b(B) = \sigma (B-B_0)$, where
$\sigma$ is the relative magnetic moment that is known for many FRs \cite{cold:chin10}.

 Introducing the normalized solution $\ket{\phi_E}$ of the open channel with
 $\ket{\Phi_P} = C \ket{\phi_E}$ and a background eigenstate $\ket{\phi_{\rm bg}}$ 
 of $\oper H_P$ with eigenenergy $E_{\rm bg}$ which is occupied for
 infinite detuning $|E-E_b|\rightarrow \infty$ one obtains the eigenenergy equation~\cite{cold:schn09a}
 \begin{equation}
  \label{eq:TC_E-E_b}
   (E-E_b)(E-E_{\rm bg}) =  
     \frac{\bra{\phi_{\rm bg}}\oper W\ket{\phi_b} \bra{\phi_b}\oper W^\dagger \ket{\phi_E}}
          { \braket{\phi_{\rm bg}}{\phi_E}}\,.
  \end{equation}

 In order to find simplified expressions for $\bra{\phi_{\rm bg}}\oper W\ket{\phi_b}$, 
 $\bra{\phi_b}\oper W^\dagger \ket{\phi_E}$, and $\braket{\phi_{\rm bg}}{\phi_E}$ one assumes that
 the interaction is only relevant in some small range $r<r_{\rm int}$ much smaller than the
 extension of the trap $r_{\rm trap}$.
 The extension of the harmonic trap is specified by the harmonic trap length $\aho = \sqrt{\hbar/(\mu\omega)}$.
 Denoting the long-range behavior of $\phi_E(r)$ 
 by $\tilde\phi_E(r)$ one finds
 \begin{equation}
 \label{eq:PhiE}
 \tilde\phi_E(r) \equiv \lim_{r\rightarrow\infty}\phi_E(r) = A_\nu D_\nu(\rho),
 \end{equation} 
 where $D_\nu(\rho)$ is the parabolic cylinder function,
 $\rho=\sqrt{2} r/a_{\rm ho}$, $\nu=E/(\hbar\omega)-1/2$, and $A_\nu$ is a normalization constant.

 For $r \ll a_{\rm ho}$ the linear approximation of $D_\nu(\rho)$ yields~\cite{cold:abra65}
 \begin{equation}
 \label{eq:PhiShortRange}
 \begin{split}  
 \tilde\phi_E(r) &= \tilde\phi_E(0) + r \tilde\phi'_E(0) + \mathcal O(r^2) \\
                 &= \tilde\phi_E(0)\left(1-  \frac{r}{a_{\rm ho} f(E)}\right) + \mathcal O(r^2)
 \end{split}
 \end{equation}
 with
\begin{equation}
 f(E) = \frac{\Gamma\left(\frac14-\frac{E}{2\hbar\omega}\right)}{2\Gamma\left(\frac34-\frac{E}{2\hbar\omega}\right)},
\end{equation}  
 where $\Gamma(x)$ is the Gamma function.
 In the range  $r_{\rm int} \ll r\ll \aho$ the radial wavefunction 
 with scattering length $\asc$ has the form $\phi_E(r) \propto 1 - r/\asc$.
 Hence, one can directly determine the scattering length of the radial wave function with energy $E$
 from Eq.~\eqref{eq:PhiShortRange}. This yields
 \begin{equation}
  \asc = - \frac{\tilde \phi_E(0)}{\tilde \phi'_E(0)}= a_{\rm ho} f(E)\,,
 \end{equation}
 which is equivalent to the result in \cite{cold:busc98}.

 In the spirit of a Taylor expansion we parametrize $\bra{\phi_E}\oper W\ket{\phi_b}$ by a 
 linear combination 
 \begin{equation}
 \begin{split}
   \label{eq:ShortRangeApp}
 \bra{\phi_b}\oper W^\dagger \ket{\phi_E} &= \gamma\, \tilde \phi_E(0) + \beta\, \tilde \phi_E'(0)\\
  &= \gamma\, \tilde \phi_E(0)\left(1 + \beta\, \frac{\tilde \phi_E'(0)}{\gamma\, \tilde \phi_E(0)}\right)\\
  &= \gamma\, \tilde \phi_E(0)\left(1 - \frac{a^*}{\asc}\right)
 \end{split}
 \end{equation}
 with $a^* = \beta/\gamma$.
 Be $\psi_b(r) = \phi_b(r)/(\sqrt{4\pi} r)$ the wave function describing the RBS then
 the expansion~\eqref{eq:ShortRangeApp} can be interpreted as approximating the coupling 
 to the bound state by $W(r)\psi_b(r) \approx \sqrt{4\pi}\gamma(r- a^*)\delta(\vec r\,)$.
 For the long-range behavior of the wavefunction $\psi_E(r)$, i.e.\ 
 $\lim_{r\rightarrow\infty} \psi_E(r) = \tilde \psi_E(r) = \tilde \phi_E(r)/(\sqrt{4\pi} r)$
 one finds
 \begin{equation}
 \begin{split}
  \gamma\, \tilde \phi_E(0) & + \beta\, \tilde \phi_E'(0) \\
 &= \int dr \tilde \phi_E(r) \left[\gamma \delta(r) +\beta \delta'(r)\right]\\
 &= \int dr \sqrt{4\pi} \tilde\psi_E(r) \left[\gamma r  - \beta \right]\delta(r)\\
 &= \int r^2 dr\, d\Omega\, \tilde\psi_E(r) \sqrt{4\pi} \gamma \left[r  - a^* \right]\delta(\vec r).
 \end{split}
\end{equation}
 Here one uses $r \delta'(r) = - \delta(r)$ and $\delta(r)=4\pi r^2\delta(\vec r)$.
 Although only two parameters are used, the parametrization of the coupling is already quite general
 since higher order couplings like those proportional to $r^2\delta(\vec r)$ automatically vanish.
 Within the approximation of a constant RBS $\gamma$ and $a^*$ must be constant.
 In reality, however, $\bra{\phi_b}\oper W^\dagger\ket{\phi_E}$ depends on the nodal structure of 
 the RBS and the open channel that are both not constant
 for a varying magnetic field. A comparison with complete coupled-channel calculations shows that
 it suffices to introduce a background coupling strength $\gamma_{\rm bg}$ for the parametrization
 of $\bra{\phi_{\rm bg}}\oper W\ket{\phi_b}$ to account for slight variations
 of the nodal structure~\cite{cold:schn11}. Since the difference between $\gamma$ and $\gamma_{\rm bg}$ 
 is only relevant for the RBS admixture but not for the eigenenergies of the system, we can ignore it 
 for our purposes. Following the reasoning given in \cite{cold:schn11}, 
 the short-range approximation \eqref{eq:ShortRangeApp} then gives 
 \begin{equation}
 \label{eq:eigenenergies}
       E-E_b = \frac{2 \gamma^2}{\aho \hbar\omega} 
                 \frac{\left(f(E)-\frac{a^*}{a_{\rm ho}}\right)\left(f(E_{\rm bg})-\frac{a^*}{a_{\rm ho}}\right)}
                      {f(E)-f(E_{\rm bg})}\,.
 \end{equation}
 The solutions of this equation determine the eigenenergies. One can rewrite 
 this equation in the form of a matching condition: The scattering length $\asc(E,E_b)$
 due to the short-range coupling to the RBS  must be equal to the product $\aho f(E)$ that 
is equal to the scattering length  
 of the long-range wavefunction $\tilde\phi_E(r)$. This yields
 \begin{equation}
 \label{eq:a_E_trap}
 \aho f(E) = \asc(E,E_{\rm res}) = a_{\rm bg}(E)\left(1-\frac{\Delta E}{E_{\rm res} - E}\right)\,.
 \end{equation}
 The right-hand side of the Eq.~\eqref{eq:a_E_trap} describes the energy dependence of the scattering length
 with background scattering length $\abg$ and resonance width
 \begin{equation}
 \label{eq:DeltaE}
  \Delta E= \frac{2 \gamma^2 \mu a_{\rm bg}}{\hbar^2}\left(1-\frac{a^*}{a_{\rm bg}}\right)^2\,.
 \end{equation}
 The resonance energy $E_{\rm res} = E_b + \delta E$ is shifted from the bound state energy $E_b$ 
 by the resonance detuning
 \begin{equation}
 \label{eq:deltaE}
  \delta E = \frac{a_{\rm bg} \Delta E}{a_{\rm bg}-a^*}\,.
 \end{equation}
 
 In the limit $E\rightarrow 0$ the ratio of the resonance detuning and the resonance width is given as
 $\delta E/\Delta E = a_0/(a_0-a^*)$, where $a_0 = \lim_{E\rightarrow 0}\abg$ is the zero-energy 
 background scattering length.
 Comparing this with the same ratio derived on the basis of a multi-channel quantum defect theory 
 for $E\rightarrow 0$ \cite{cold:gora04} allows for removing one free parameter $a^*$. One finds
 \begin{equation}
 \label{eq:aStar}
   a^* = \overline a \left( 1 + \frac{\overline a}{\overline a - a_0}\right)\,,
 \end{equation}
 where the mean scattering length $\overline a$ is determined by the $C_6$ coefficient of the van der Waals 
 interaction \cite{cold:grib93}. 
 Using Eqs.~\eqref{eq:DeltaE} and \eqref{eq:aStar}, the remaining parameter $\gamma$ can be directly related 
 to the resonance width $\Delta E$.
 
 The function $\aho f(E)$ which describes the scattering length of the wave function $\tilde \phi_E(r)$ is also known
 for anisotropic traps with $\omega_y=\omega_z=\eta \omega_x$. In this case the scattering length is given as
 $\asc = -\sqrt{\pi} d /\mathcal F(u,\eta)$ (with $d,u=u(E),$ and $\mathcal F$ defined in \cite{cold:idzi06}) such that
 the eigenenergy relation 
 \begin{equation}
 \label{eq:a_E_trap_aniso}
 -\frac{\sqrt{\pi} d}{\mathcal F(u,\eta)} = a_{\rm bg}(E)\left(1-\frac{\Delta E}{E_{\rm res} - E}\right)\,
 \end{equation}
 holds.   
    
One generally distinguishes between narrow and broad FRs \cite{cold:chin10}.
In the case of a broad FR the coupling strength to the bound state is 
relatively large such that it is 
admixed to unbound states in a large energy domain. 
If, as usual, the background scattering length $\abg$ is small compared to the trap length $\aho$, 
the ratio of the RBS admixture $|A|^2$ to the open-channel admixture 
$|C|^2$ for states above the first trap state is on the order of $\aho\hbar\omega/(\abg \Delta E)$ 
such that the RBS admixture can be neglected if $\abg \Delta E \gg \aho\hbar\omega$ \cite{cold:schn11}.
Furthermore, also the energy dependence of the scattering length becomes negligible if
$\abg \Delta E \gg \aho\hbar\omega$ (see Fig.~\ref{fig:SketchBroadVsNarrow}).
Therefore, all details of the atomic interaction apart from the value of scattering length for $E\rightarrow 0$ 
are irrelevant. This situation is called universal.
On the other hand, for narrow FRs the bound state couples only to a narrow energy range of
scattering states or respectively to that unbound trap state which is 
in resonance. As shown in Fig.~\ref{fig:SketchBroadVsNarrow} in a harmonic trap this leads to narrow 
avoided crossings in the energy spectrum with an energy splitting on the order of 
$\sqrt{\abg \Delta E/(\aho \hbar\omega)}\hbar\omega$ \cite{cold:schn11}. 
At the resonance the bound state is strongly admixed and the energy dependence of the 
scattering length cannot be neglected. 

\section{Feshbach resonance in an optical lattice}
\label{sec:FRinOL}

In order to avoid unnecessary complexity, in the following an OL is considered, in which two directions 
of movement are effectively frozen out by using strong harmonic confinement.
Nevertheless, the following discussions can be easily extended to 2D and 3D lattices.

A particle of mass $m$ in such an OL of depth $V_{\rm L}$ and
periodicity $d=\pi/k_0$ in the spacial direction $x$ and transversal
harmonic confinement with frequency $\omega_{\rm t}$ in $y$ and $z$ direction is described by the Hamiltonian
\begin{equation}
 \oper{\mathcal H}_m = \frac{\hat p^2}{2m} + V_{\rm L} \sin^2(k_0 \hat x) + \frac12 m \omega_{\rm t}^2 (\hat y^2 + \hat z^2) \, .
\end{equation}
Eigensolutions of this Hamiltonian with quasi momentum $k$ are given by 
\begin{equation}
 \label{eq:bloch_basis}
 \Phi_{k,n,M_y,M_z}(x,y,z)=e^{i k x}\phi_{n,k}(x)\,H_{M_y}(y)\,H_{M_z}(z) 
\end{equation} 
where $\phi_{n,k}$ are analytically known Bloch solutions with band index $n=1,2,3,\dots$ and 
quasi momentum $k$ of the periodic lattice. $H_M$ is the $M$-th solution of the one-dimensional
harmonic oscillator with transversal frequency $\omega_{\rm t}$.

In order to describe more than one particle in an OL, interactions
have to be taken into account. 
Since neutral atoms interact only on short distances it is convenient to transform 
the basis \eqref{eq:bloch_basis} into localized functions. This is done by
the usual transformation to Wannier functions \cite{cold:kohn59}
\begin{equation}
\label{eq:wannier_basis}
 W_{i,n,M_y,M_z}(x,y,z)=\mathcal W_{i,n}(x)\,H_{M_y}(y)\,H_{M_z}(z)\,.
\end{equation}
Here, $\mathcal W_{i,n}$ denotes the Wannier function localized at lattice site $i$ and band $n$.

Due to the anharmonicity of the OL the relative-motion (\reli)\ coordinates 
$\vec r = (x,y,z)^T = \vec r_1 - \vec r_2$ and center-of-mass (\cmi) coordinates 
$\vec R =  (X,Y,Z)^T = (\vec r_1+ \vec r_2)/2$ are coupled.
Therefore, the Eqs.~\eqref{eq:two_channel_1} and \eqref{eq:two_channel_2}
for \rel motion have to be extended to include also the \cm energies of the two 
atoms and the resonant molecular state.
To this end $\Psi_P(\vec r_1,\vec r_2\,)$ shall describe the wave function of the 
two atoms in the open channel with kinetic and potential energies 
$\mathcal H_m(\vec r_1) + \mathcal H_m(\vec r_2)$ interacting via a short-range potential
$V(r)$.
The open channel is coupled by some real-valued short-range coupling $W(r)$ to 
the closed-channel wave function $\Psi_Q(\vec R, \vec r)$.
One assumes that the RBS in \rel motion has an extension small enough not to
probe the external trapping potential. Therefore, the closed-channel wave function can be written
as a product state $\Psi_Q(\vec R, \vec r) = \psi_b(\vec r\,)\Psi_{\cmi}(\vec R)$ 
of the RBS $\psi_b(\vec r\,)$ with binding energy  $E_b$, which is equal to the 
one introduced in Sec.~\ref{sec:FRinHarmTrap}, and the \cm wave function $\Psi_{\cmi}(\vec R)$
that experiences the kinetic and potential energy of a particle of mass $2m$, $\mathcal H_{2m}(\vec R\,)$.

Consequently, two atoms in an OL at a Feshbach resonance are described by the
coupled equations
\begin{equation}
\label{eq:TCinRealSpace}
\begin{split}
&\begin{pmatrix}
\mathcal H_m(\vec r_1) + \mathcal H_m(\vec r_2) + V(r) &  W(r)\\
W(r) & \mathcal H_{2m}(\vec R\,) + E_b 
\end{pmatrix}\times\\
&\qquad\times \begin{pmatrix}
 \psi_{\rm P}(\vec r_1,\vec r_2\,)\\
 \psi_b(\vec r)\Psi_{\cmi}(\vec R)
\end{pmatrix}
= E \begin{pmatrix}
 \psi_{\rm P}(\vec r_1,\vec r_2\,)\\
 \psi_b(\vec r)\Psi_{\cmi}(\vec R)
\end{pmatrix}\,.
\end{split}
\end{equation}

As is usually done for Hubbard models the Hamiltonian is reformulated in the basis of Wannier functions of the OL.
However, in order to include effects of higher Bloch bands and their couplings due to the presence of the
RBS the basis is not restricted to the first Bloch band.
In the following the simplification of strong transversal confinement is considered, i.e.\
the ultracold atoms only occupy the ground state of transversal motion. Let $a_{i,n}^\dagger$ ($a_{i,n}$) 
be the creation (annihilation) operator of an atom with Wannier function $w_{i,n} \equiv W_{i,n,0,0}$ 
and $b_{i,n}^\dagger$ ($b_{i,n}$) the creation (annihilation) operator of the RBS with \cm Wannier function $\tilde 
w_{i,n} \equiv \tilde W_{i,n,0,0}$ \footnote{The Wannier functions of atoms and molecules differ due to their different 
mass.}. The Hamiltonian in second quantization that is equivalent to the coupled equations 
\eqref{eq:TCinRealSpace} expanded in the Wannier basis is given as
\begin{equation}
\label{eq:TCSecondQuantized}
\begin{split}
\oper H &= \sum_{i,j}\sum_{n,m} \QMa{w_{i,n}}{\oper{\mathcal{H}}_m}{w_{j,m}} a_{i,n}^\dagger a_{j,m}  \\
& +\frac12\sum_{i,j,k,l}\sum_{n,m,p,q} \QMa{w_{i,n} w_{j,m}}{\oper V}{w_{k,p} w_{l,q}} a_{i,n}^\dagger a_{j,m}^\dagger a_{k,p} a_{l,q} 
\\
& +\sum_{i,j}\sum_{n,m} \left(\QMa{\tilde w_{i,n}}{\oper{\mathcal{H}}_{2m}}{\tilde w_{j,m}} + E_b \right) b_{i,n}^\dagger b_{j,m} \\
& +\frac{1}{\sqrt{2}} \sum_{i,j,k}\sum_{n,m,p} \QMa{w_{i,n} w_{j,m}}{\oper W}{\tilde w_{k,p}\; \psi_b}\left( a_{i,n}^\dagger a_{j,m}^\dagger b_{k,p}  + h.c. \right)\,.
\end{split}
\end{equation}
Note the factor $1/\sqrt{2}$ before the atom-molecule coupling, which
has to be included to ensure that the matrix elements of
the Hamiltonian are equal in first and second quantization \cite{cold:timm99}.

We want to emphasize that Eq.~\eqref{eq:TCinRealSpace} and thus the second quantized Hamiltonian
\eqref{eq:TCSecondQuantized} are only valid if not more than two atoms interact.
For more atoms important effects such as losses or the appearance of Efimov states cannot be correctly reproduced.

The following simplifications and approximations are introduced:
\begin{enumerate}
 \item The Hamiltonians $\mathcal H_m$ and $\mathcal H_{2m}$ do not couple different Bloch bands, 
 since the Bloch functions $w_{i,n}$ and $\tilde w_{i,n}$ are eigenstates of $\mathcal H_m$ and 
 $\mathcal H_{2m}$, respectively. 
 For example, for $\mathcal H_m$ holds $\QMa{w_{i,n}}{\oper{\mathcal{H}}_m}{w_{j,k}} = 
  \QMa{w_{i,n}}{\oper{\mathcal{H}}_m}{w_{j,n}}\delta_{n k}$.
%
\item Only next-neighbor coupling is considered, i.e.
\[
\begin{split}
   &\sum_{i,j}\sum_{n}         \QMa{w_{i,n}}{\oper{\mathcal{H}}_m}{w_{j,n}} a_{i,n}^\dagger a_{j,n} \\
  &+ \sum_{i,j}\sum_{n,m} \left(\QMa{\tilde w_{i,n}}{\oper{\mathcal{H}}_{2 m}}{\tilde w_{j,m}} + E_b \right) b_{i,n}^\dagger b_{j,m} \\ 
 &\approx  
 \sum_{i}\sum_{n} \epsilon_{n} a_{i,n}^\dagger a_{i,n} - \sum_{\left<i,j\right>}\sum_{n} J_{n} a_{i,n}^\dagger a_{j,n} \\
 & + \sum_{i}\sum_{n} \left(\mathcal E_{n}+E_b\right) b_{i,n}^\dagger b_{i,n} - \sum_{\left<i,j\right>}\sum_{n} \mathcal J_{n} b_{i,n}^\dagger b_{j,n}
\end{split}
\]
where $\left<\cdots\right>$ denotes summation over nearest-neighbor lattice sites,
$\epsilon_{n}=\QMa{w_{1,n}}{\oper{\mathcal{H}}_m}{w_{1,n}}$,
$\mathcal E_{n} = \QMa{\tilde w_{1,n}}{\oper{\mathcal{H}}_{2m}}{\tilde w_{1,n}}$,
$J_{n} = -\QMa{w_{1,n}}{\oper{\mathcal{H}}_m}{\tilde w_{2,n}}$, and
$\mathcal J_{n} = -\QMa{\tilde w_{1,n}}{\oper{\mathcal{H}}_{2 m}}{w_{2,n}}$.
\item The interaction potential is replaced by the Fermi-Huang pseudo potential $V(r) \rightarrow 
\frac{4 \pi \hbar^2 \abg}{m}\delta(\vec r)\frac{\partial}{\partial r}r$ that reproduces the same 
background scattering length $\abg$ as the full open-channel interaction potential. 
For small background scattering length only onsite-interaction is taken into account, i.e.
\[
\begin{split}
&\sum_{i,j,k,l}\sum_{n,m,p,q} \QMa{w_{i,n} w_{j,m}}{\oper V}{w_{k,p} w_{l,q}} a_{i,n}^\dagger a_{j,m}^\dagger a_{k,p} a_{l,q} \\
&\qquad\approx \sum_{i}\sum_{n,m,p,q} U_{n,m,p,q} a_{i,n}^\dagger a_{i,m}^\dagger a_{i,p} a_{i,q} 
\end{split}
\]
with $U_{n,m,p,q} = \QMa{w_{1,n} w_{1,m}}{\oper V}{w_{1,p} w_{1,q}}
   = \frac{4 \pi \hbar^2 \abg}{m}  \int d x\; d y\; d z\; w_{0,n}\; w_{0,m}\; w_{0,p}\; w_{0,q}$.
\item The coupling to the molecule happens only at short distances, i.e.\ on the length scale of the lattice and the transverse harmonic 
confinement, thus one can replace 
\begin{equation}
\label{eq:coupling _to_bound_state_BH}
W(\vec r\,)\psi_b(\vec r\,)\rightarrow g \delta(\vec r\,)\,,
\end{equation} 
where the coupling strength $g$ has to be adapted to match the behavior of the system under consideration. 
Including only next-neighbor coupling leads to the simplification
\[
\begin{split}
\sum_{i,j,k}\sum_{n,l,p}& \QMa{w_{i,n} w_{j,l}}{\oper W}{\tilde w_{k,p}\; \psi_b}\left( a_{i,n}^\dagger a_{j,l}^\dagger b_{k,p}  + h.c. \right) \\
&\approx 
\sum_{\left<i,j,k\right>}\sum_{n,l,p} g_{n, m, p}^{(i-k,j-k)} \left( a_{i,n}^\dagger a_{j,l}^\dagger b_{k,p}  + h.c. \right)\,,
\end{split}
\]
with 
\begin{equation}
\label{eq:coupling energy}
g_{n, l, p}^{(i,j)} = g \int dx \; d y\;  d z\;  w_{i,n}\; w_{j,l}\; \tilde w_{0,p}\,.
\end{equation}
Due to the symmetry of the Wannier functions the onsite coupling obeys the selection rule
\[g_{n, l, p}^{(0,0)} = 0 \quad\text{for}\quad n+l+p \quad\text{even.}\]
\end{enumerate}

Employing the above simplifications and approximations the BH Hamiltonian
reduced to the first $N$ Bloch bands is given as
\begin{equation}
\label{eq:Hamiltonian_BH}
\begin{split}
\oper H_{\rm BH} =&
   \sum_{i}\sum_{n=1}^N \epsilon_{n}\; a_{i,n}^\dagger a_{i,n} -
   \sum_{\left<i,j\right>}\sum_{n=1}^N J_{n}\; a_{i,n}^\dagger a_{j,n}\\
&+ \frac12\sum_{i}\sum_{n,l,p,q=1}^N U_{n,l,p,q}\; a_{i,n}^\dagger a_{i,l}^\dagger a_{i,p} a_{i,q} \\
&+ \sum_{i}\sum_{n=1}^N \left(\mathcal E_{n}+E_b\right)\; b_{i,n}^\dagger b_{i,n} -
    \sum_{\left<i,j\right>}\sum_{n=1}^N \mathcal J_{n}\; b_{i,n}^\dagger b_{j,n}\\
&+ \frac{1}{\sqrt{2}}\sum_{\left<i,j,k\right>}\sum_{n,l,p=1}^N g_{n, m, p}^{(i-k,j-k)} \left( a_{i,n}^\dagger a_{j,l}^\dagger b_{k,p}  + h.c. \right)\,.
\end{split}
\end{equation}

\section{Problem of representing a delta-like coupling within the Bose-Hubbard model}
\label{sec:Problem}

The coupling of the open channel to the bound state as described by Eq.~\eqref{eq:coupling _to_bound_state_BH}
seems to be a crude approximation. Indeed, as discussed in Sec.~\ref{sec:FRinHarmTrap}, a more general 
form of a short-range coupling to the
bound state is of the form $W(\vec r\,)\psi_b(\vec r\,) = \sqrt{4\pi}\gamma(r-a^*)\delta(\vec r\,)$.
While one can associate $g$ with $\sqrt{4\pi}\gamma a^*$ the coupling $\sqrt{4\pi}\gamma r\delta(\vec r\,)$
automatically vanishes for the chosen single-atom basis states. In fact it vanishes for any basis
that conforms to a scattering length $\asc=0$.
Hence, the presented BH model can only conform to a FR with $\gamma=0$, $a^*\rightarrow \infty$,
and $\gamma a^* = const.$
This is only the case for $a_0 = \overline a$ [see Eq.~\eqref{eq:aStar}] and results
according to Eqs.~\eqref{eq:DeltaE} and \eqref{eq:deltaE} in a resonance width 
$\Delta E = \frac{2 \mu }{\hbar^2}\frac{(\gamma a^*)^2}{a_{\rm bg}} = \frac{2 \mu g^2}{4\pi \hbar^2 a_{\rm bg}}$ 
and a resonance detuning $\delta E = 0$.
For FRs with $\gamma\neq 0$ one can easily account for the altered resonance parameters by introducing 
an effective coupling strength and an effective bound-state energy,
\begin{eqnarray}
g &\rightarrow& g_{\rm eff} = \sqrt{\frac{4\pi \hbar^2 a_{\rm bg} \Delta E}{2\mu}}\\
E_b &\rightarrow& E_{b, \rm eff} = E_{\rm res} = E_b + \delta E\,
\end{eqnarray}
that lead to the correct resonance width $\Delta E$ and resonance energy $E_{\rm res}$.
In the following the index ``eff'' will be suppressed keeping however in mind that 
$g$ and $E_b$ are not equivalent to the physical coupling strength and the physical energy
of the RBS.

\begin{figure}[ht]
 \centering
 \includegraphics[width=0.95\linewidth]{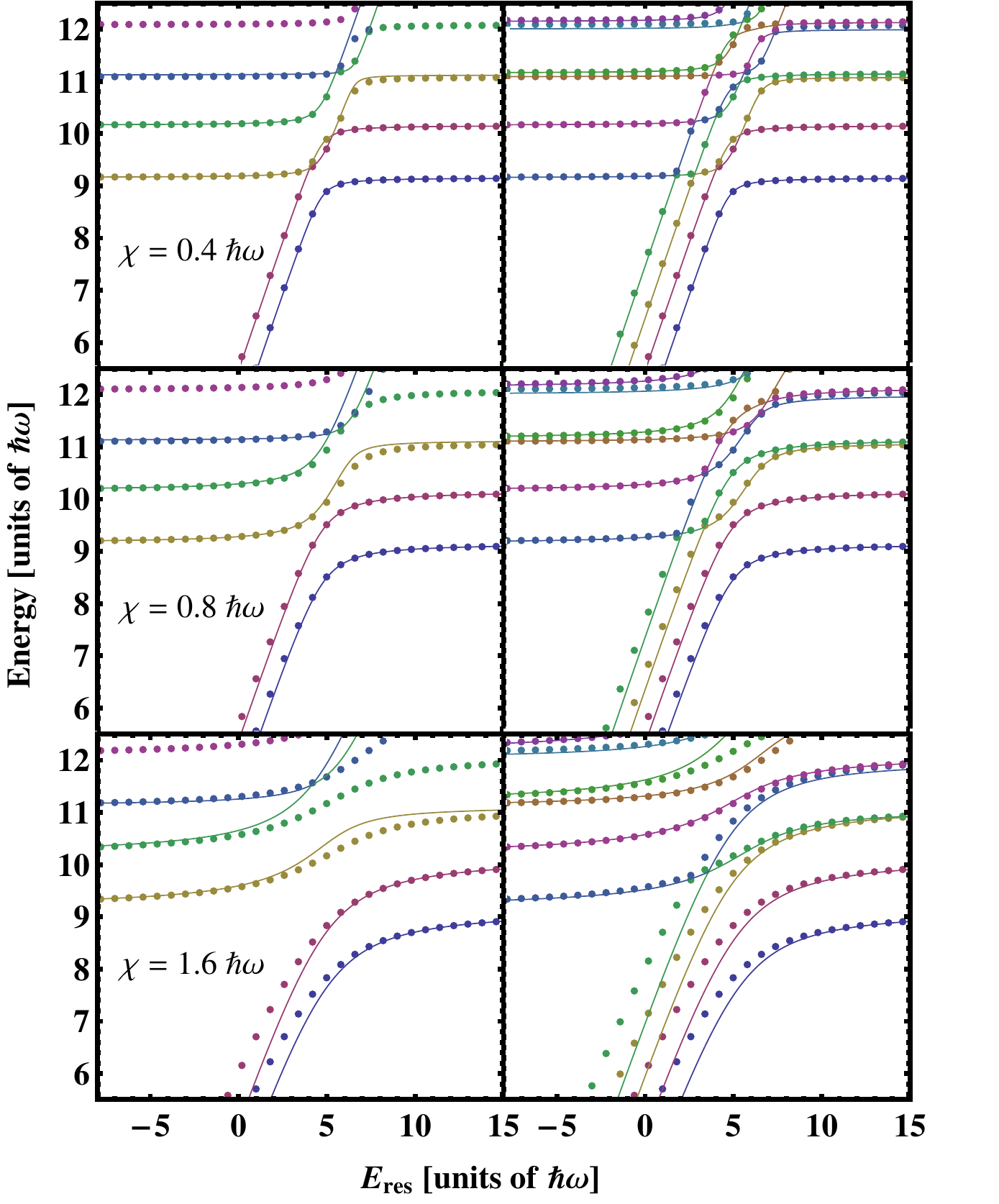}
 \caption{(Color online). Energy spectrum as a function of the resonance energy $E_{\rm res}$ for 
  $\eta=4$, $\abg = 0.04 \aho$ and from top to bottom $\Delta E = (1, 4, 16) \hbar\omega$. 
  This results in the coupling energies $\chi$ [see Eqs.~\eqref{eq:coupling _to_bound_state_BH} and 
  \eqref{eq:chi}] given in the graphs.
  The analytic eigenenergies (dots) obtained by Eq.~\eqref{eq:a_E_trap_aniso} are compared to the 
  eigenenergies of the BH model (lines) for two Bloch 
  bands (left column) and four Bloch bands (right column) included.
  Including only two Bloch bands (left column) the analytic eigenstate with energy $\approx 12 \hbar\omega$ is not reproduced
  by the BH model.}
 \label{fig:SpectrumHarmonic}
\end{figure}

In Fig.~\ref{fig:SpectrumHarmonic} the energy spectra in an anisotropic harmonic trap  of several FRs 
of different widths are compared to the corresponding result of the effective BH model.
The trapping frequencies are $\omega_y=\omega_z=\eta \omega$, with $\eta = 4$ and $\omega$ the trapping
frequency in $x$ direction.
In the harmonic trap the Wannier functions of the BH model are replaced by harmonic-oscillator eigenfunctions.
On the left side two Bloch bands are included and the RBS appears in two different \cm states
while the unbound atoms can occupy three different trap states [(i) both atoms in the first band at 
$9\hbar\omega$, (ii) one atom in the first and one in the second band at $10 \hbar\omega$ and 
(iii) two atoms in the second band $11 \hbar\omega$].
On the right side four Bloch bands are included with correspondingly more molecular states and trap states.

As a measure for the coupling strength the energy
\begin{equation}
\label{eq:chi}
\chi = g_{1, 1, 1}^{(0,0)}
\end{equation}
is introduced [see Eq.~\eqref{eq:coupling energy}]. The avoided crossing between the lowest bound state and the first 
trap state has a splitting energy of $\approx 2\chi$.

For a relatively narrow FR with an effective coupling strength $\chi=0.4\hbar\omega$ the agreement between the BH model 
and the analytic result is very good independently of the number of Bloch bands included.
For the broader FRs with $\chi=0.8\hbar\omega$ and $\chi=1.6\hbar\omega$ one can make two observations: 
(i) Trap states (i.e.\ states above the bound state threshold of $9\hbar \omega$) quickly approach to the analytic results
    for an increasing number of Bloch bands.
(ii) The disagreement between analytic and BH results of the bound states does not decrease with the number of Bloch bands.

Obviously, the variational principle does not hold for the bound state 
as an insufficient basis leads to an energy \emph{lower} than the correct bound state energy. 
Moreover, by increasing the basis the already incorrect bound-state energy becomes even lower and the disagreement 
to the correct result increases. Though less severe, the same problem also appears for trap states. 
For example, the first trap state in the last row in 
Fig.~\ref{fig:SpectrumHarmonic} lies below the correct energy if four Bloch bands are included.

\begin{figure}[ht]
 \centering
 \includegraphics[width=0.85\linewidth]{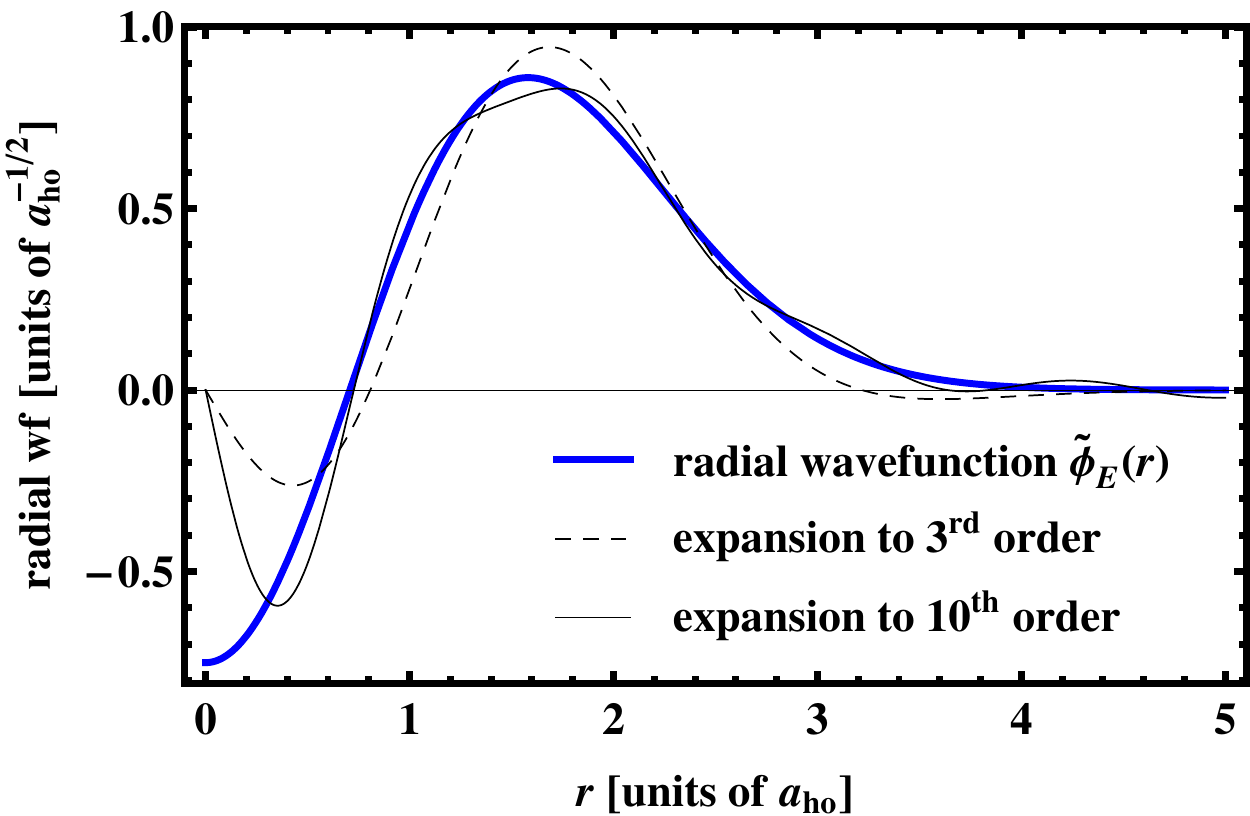}
 \caption{
 (Color online). The radial wave function $\tilde\phi_E(r)$ in a spherical harmonic trap
 introduced in Eq.~\eqref{eq:PhiE} is compared for the \rel energy $E=2.5\hbar\omega$ to its expansion 
 $\phi_{\rm exp} =\sum_{n=0}^{N-1} \langle \phi_n | \phi_E \rangle \phi_{n}(r)$ to different orders $N$, 
 where $\phi_{n}$ is the radial wave function of the non-interacting system with radial momentum $l=0$ 
 and energy $(2n+\frac{3}{2})\hbar\omega$. Since all non-interacting radial basis functions are zero for $r=0$
 the expansion cannot reproduce the behavior of $\tilde\phi_E(r)$ for $r\rightarrow 0$.
 This is important, since the coupling to the bound state is proportional to $\tilde\phi_E'(0)$ or $\phi_{\rm exp}'(0)$, respectively.  
 }
 \label{fig:Expansion}
\end{figure}

The reason for this insufficiency of the basis to conform to the behavior of a delta-like coupling is related to
the problem of a missing coupling of the form $\sqrt{4\pi}\gamma r\delta(\vec r\,)$: the two-particle basis states 
are $\asc=0$ wave functions. 
However, $\asc=0$ basis functions can represent the full wave function only for $r>0$ but not
for $r \rightarrow 0$ (see Fig.~\ref{fig:Expansion}). While for ordinary interaction potentials 
the value of the wave function at $r=0$ is irrelevant, for zero-range potentials it is decisive.
The problem is especially severe for the open-channel bound state, which appears for positive scattering lengths.
For $E\rightarrow -\infty$ one has $|\tilde\phi_E(0)| \propto (-E)^{1/4}$ making 
its representation by $\asc=0$ basis functions for decreasing energy more and more problematic.

For weak coupling the problem is less severe as eigenstates that differ significantly 
from the background trap states are predominantly bound states with different \cm
excitations, which are well reproduced by the BH model. 
For strong coupling, however, the bound state is admixed to many states in the spectrum 
(see Sec.~\ref{sec:FRinHarmTrap}). Since the bound-state
admixture for a certain eigenstate is thus lower, a good representation of the open-channel wavefunction is important
also for large scattering lengths.

The described problem does not only arise when using non-interacting $\asc=0$ basis states. For any finite expansion of the 
radial wave function $\phi_{\rm exp}(r) = \sum c_n \phi_n(r)$ in a superposition of basis functions with a specific
scattering length $a_b$ [i.e. $a_b = -\phi_n(0)/\phi_n'(0)$] the scattering length of the expansion yields 
\begin{equation}
a_{\rm exp} = -\frac{\phi_{\rm exp}(0)}{\phi_{\rm exp}'(0)} 
            = \frac{\sum c_n a_b \phi_n'(0)}{\sum c_n \phi_n'(0)}
            = a_b\,.
\end{equation}
Hence, the wave function $\phi_{\rm exp}(r)$ cannot adapt to a change of the scattering length induced 
by a short-range coupling.
Especially, since the scattering length at a FR is energy dependent these expansions cannot reproduce 
the correct eigenenergies and eigenstates.

\section{Dressing of coupling strength and bound-state energy}
\label{sec:dressing}

To circumvent the problem of the wrong representation of a zero-range coupling one can replace it by a finite-range coupling.
To this end one usually considers the Fourier transform of the problem and regularizes the delta-like 
interaction by introducing a high-momentum cut-off $\Lambda$. Thereupon the coupling parameter
is renormalized \cite{cold:cava99}.
Taking the limit $\Lambda\rightarrow\infty$ the finite-range coupling converges 
towards a zero-range coupling. However, for an interaction with a range of $d/N$ where $d$ is the lattice spacing
more than $N$ Bloch bands have to be included to converge the energies \cite{cold:buch12}.

Here we want to take a different approach with no need to include more Bloch bands to reproduce the 
correct bound-state energies.
Provided with the analytic solution in the harmonic trap a dressed bound state is introduced, which reproduces the
correct energy spectrum in the harmonic trap at least in the important energy range of the first Bloch band.
We use the fact that the full bound state (the combination of the closed-channel and open-channel bound state)
falls off rapidly for increasing internuclear separation. Hence, the bound state does hardly probe the anharmonic parts of 
the potential and the dressed bound state can be equally used for (anharmonic) OLs.

\begin{figure}[ht]
 \centering
 \includegraphics[width=0.9\linewidth]{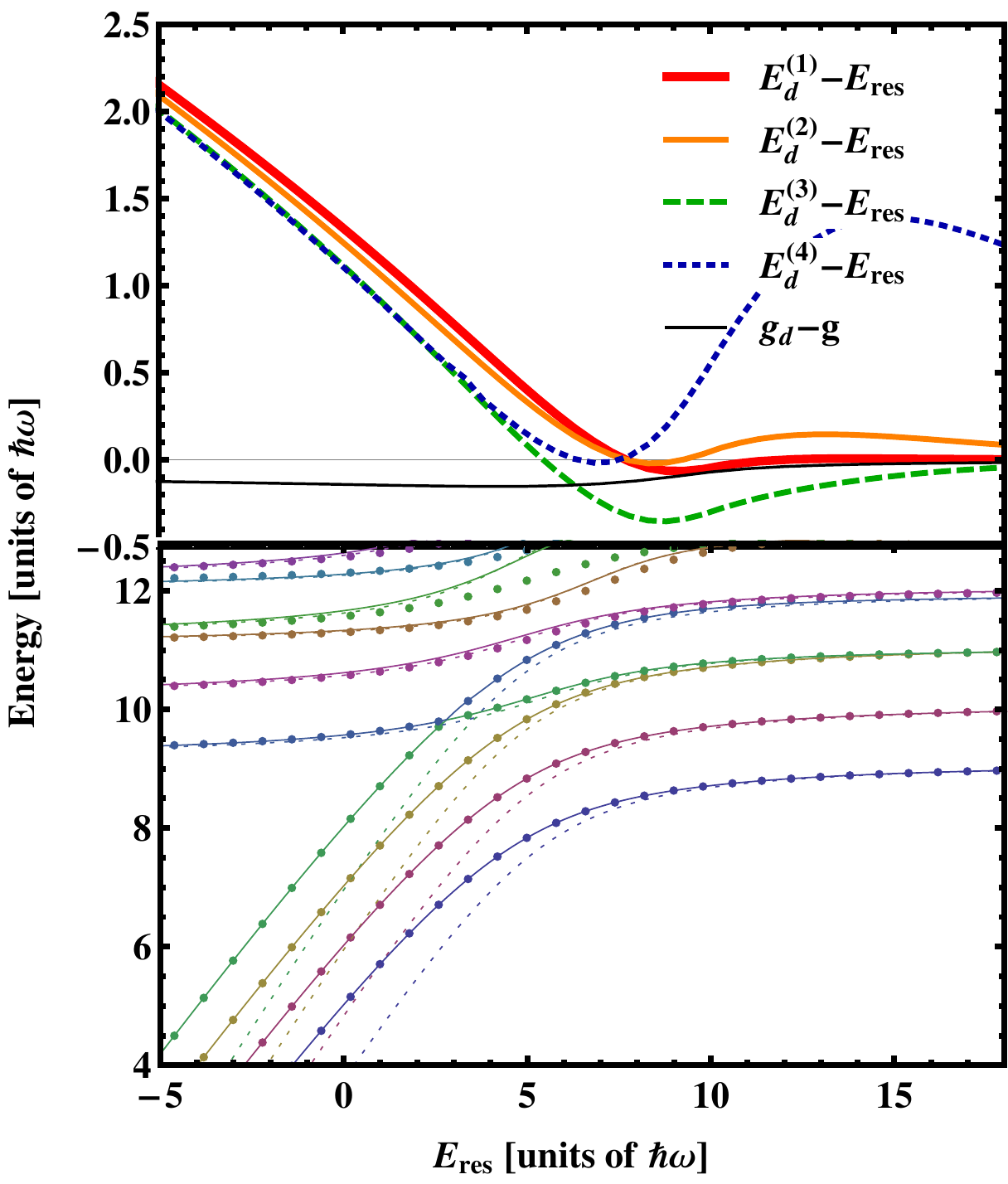}
 \caption{(Color online). Results of the dressed BH model for four Bloch bands with $\abg = 0.04\aho$ and 
 $\Delta E = 16 \hbar\omega$. {\bf Top}: Dressed bound-state energies and dressed coupling strength as a function 
 of $E_{\rm res}$. {\bf Bottom}: Comparison of the analytic energy spectrum (dots) with the energies of the dressed
 BH model (solid lines) and the undressed BH model (dotted lines). 
}
 \label{fig:CorrectedHarmonic}
\end{figure}

\begin{figure}[ht]
 \centering
 \includegraphics[width=0.9\linewidth]{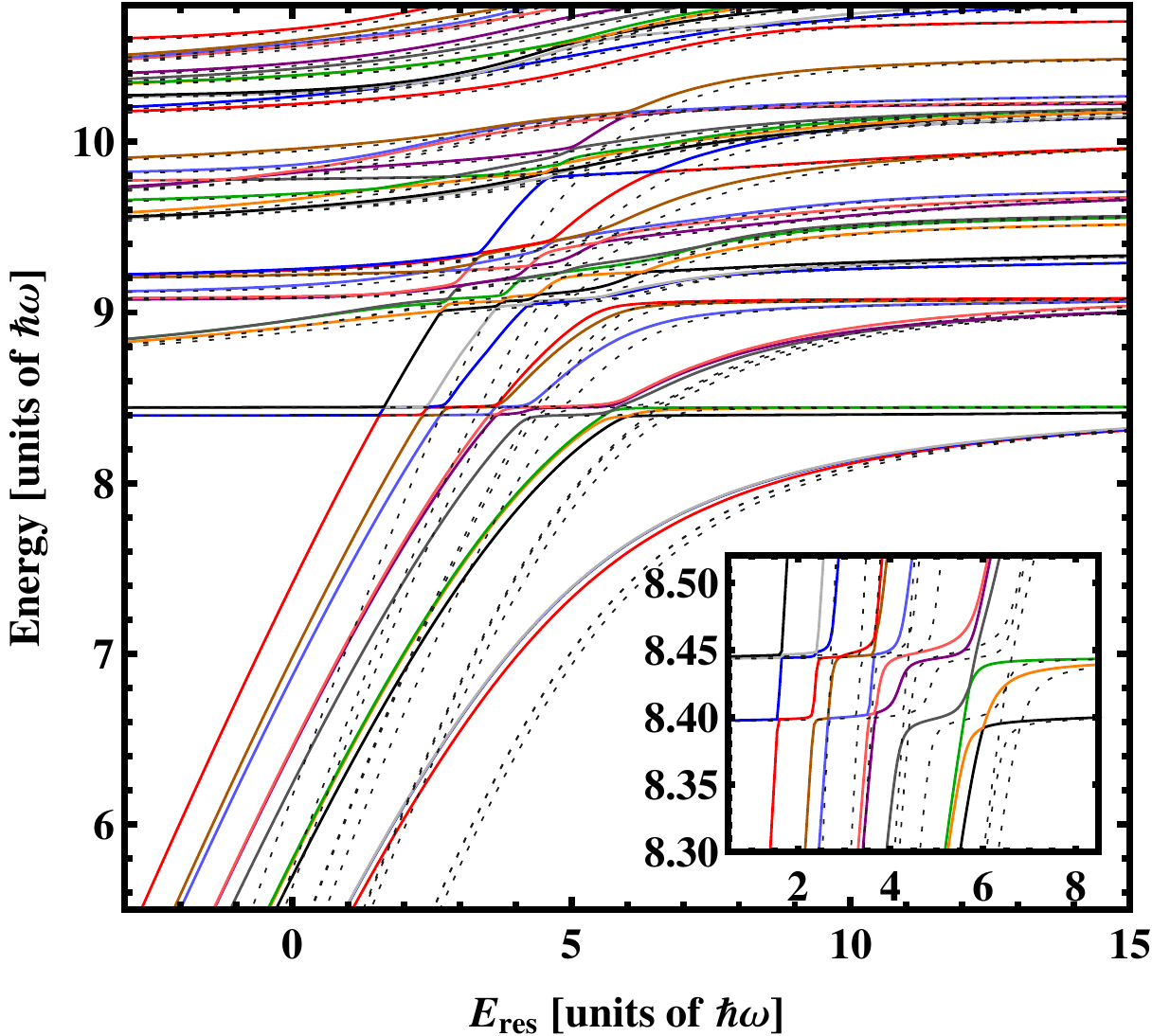}
 \caption{(Color online). Energy spectrum of two atoms in an OL with lattice depth $V_{\rm L} = 5 E_r =1.1 \hbar\omega$
 consisting of three lattice sites with periodic boundary conditions.
 Excitations in transversal direction are frozen out by choosing transversal trapping frequencies 
 $\omega_y = \omega_z = 3.8 \omega$, where $\omega$ is the frequency of the harmonic approximation of a lattice site
 in $x$ direction.
 The resonance parameters are  $\abg =85\,\text{a.u.} = 9.0\cdot 10^{-3}d$, and 
 $\Delta E= 24.2 \hbar\omega$ which corresponds to a coupling strength of $\chi = 1.66\,\hbar\omega = 1.48\,V_{\rm L}$ 
 (See also the right graph in Fig.~\ref{fig:SpectraLattice} with the same lattice parameters
 and resonance parameters).
 The comparison of the eigenenergies of the dressed BH model (solid lines) and the undressed BH model
 (dashed lines) each with four Bloch bands included, shows that again both models disagree especially for 
 the bound states, while the differences for the trap states are small.
 The inset shows a magnification of the spectrum close to the crossing of excited bound states with the lowest Bloch 
 band.
}
 \label{fig:spectrumBroadLattice}
\end{figure}

More concretely, the dressed bound state is introduced in the following way: The RBS in the first band 
(for which the \cm wave function is a Wannier function of the first band) couples predominantly to two atoms in the first
band leading to the lowest avoided crossing in the spectrum. The two corresponding eigenenergies are given by a sum of
the lowest \cm energy $E^{\rm \cmi}_1$ [$E^{\rm \cmi}_n =\hbar\omega (n-\frac{1}{2}+\eta)$] and the two lowest solutions $E_1, E_2$ of 
the \rel motion eigenenergy relation \eqref{eq:a_E_trap_aniso} which depend on the bound-state energy $E_b = E_{\rm res}$. 
In order to match the energies of this avoided crossing the
bound-state energy $E_b$ and the coupling strength $g$ are replaced by dressed parameters
$E_b \rightarrow E_d^{(1)}(E_{\rm res})$ and $g\rightarrow g_d$.
The two parameters are determined by a least square fit to the energies $E_1+E_0^{\rm \cmi}$ and $E_2+E_0^{\rm \cmi}$.

To match the energies $E_1+E_n^{\rm \cmi}$ with $n=2,3,\dots$ of bound states in higher Bloch bands, 
dressed bound-state energies $E_d^{(2)}(E_{\rm res}),E_d^{(3)}(E_{\rm res}),\dots$ are introduced, which are also determined by a least square fit.
The upper branches of the avoided crossings with bound states in higher Bloch bands lay above the first Bloch band.
Therefore, their correct representation is less 
relevant and we do not need to introduce also band-dependent dressed coupling strengths.

In Fig.~\ref{fig:CorrectedHarmonic} the dressed energies $E_d^{(1)},E_d^{(2)},E_d^{(3)},E_d^{(4)}$ and $g_d$ and the 
corresponding corrected spectrum are shown for the four-band BH model with $\abg = 0.04\aho$ and 
 $\Delta E = 16 \hbar\omega$ (same parameters as for right bottom graph in Fig.~\ref{fig:SpectrumHarmonic}).
Evidently, the dressing of the bound states becomes relevant for a resonance energy $E_{\rm res}<5\,\hbar \omega$, 
but is already visible for $E_{\rm res}<10\,\hbar \omega$ .
Since only a band-independent dressed coupling strength was introduced, the repulsive branches above the first Bloch band
with an energy above $10\hbar\omega$ are not fitted to the exact results.
Correspondingly, slight deviations between the exact energies and the dressed BH energies appear for these states, 
while the first repulsive branch is correctly reproduced.

The introduced dressed parameters can now be used to determine the energy spectrum of two atoms in an OL.
In Fig.~\ref{fig:spectrumBroadLattice} the spectrum of the dressed and undressed BH model of two atoms 
in a small OL consisting of three lattice sites are compared for a coupling energy of $\chi = 1.66\,\hbar\omega = 1.48\,V_{\rm L}$.
In contrast to the purely harmonic trap, the energies of the bound states and the trap states split due to tunneling. 
If the molecular bound states are not in resonance, i.e.\ for $E_{\rm res}<0$, the trap-state energies form 
bands of increasing widths around $8.4\hbar\omega$, $9.1\hbar\omega$, $9.8\hbar\omega$, and $10.4\hbar\omega$.
For resonance energies $E_{\rm res}>0$ the bound states cross with the trap states leading to a plethora
of avoided crossings. In the ultracold regime especially the avoided crossings with the first band are of 
relevance. These appear due to the next-neighbor coupling of the molecular state with the atomic states
\cite{cold:stec11}. As shown in the inset of
Fig.~\ref{fig:spectrumBroadLattice} the width of these avoided crossings decreases with the \cm 
excitation energy of the RBS.
The comparison between the dressed and the undressed BH model shows that also in the OL the energies disagree 
especially for the bound states, while the energy differences for the trap states are small.

\section{Non-perturbative determination of stationary and dynamical states}
\label{sec:numMethod}

In the following the results of the BH model shall be compared to non-perturbative calculations for two
atoms at a FR in an OL consisting of two lattice sites. In order to do so
an approach described in \cite{cold:gris11} is used, which allows for finding the stationary solutions of 
the two-body problem with arbitrary isotropic \emph{single-channel} interaction potentials.
On the basis of the stationary solutions the method described in \cite{cold:schn12a} is used to
determine the time-dependent wavefunction during a perturbation of the lattice potential.

Since the lattice potential couples \rel and \cm motion and the interaction couples the motion
in $x$, $y$ and $z$ direction all six coordinates of the problem are coupled. An extension to 
the coupling to an additional channel describing the \cm and \rel motion of the molecular bound state
would make the solution very cumbersome. Instead, the freedom of the choice of the interaction potential is used to 
realistically mimic a two-channel problem by a square-well interaction potential. The potential supports 
bound states that are coupled by a barrier to the scattering states. In the following it is shown that this 
potential leads to an energy dependence of the scattering length, which is in very good agreement to the one
of a two-channel description in Eq.~\eqref{eq:a_E_trap}.
This is already sufficient to realistically mimic a FR since, as shown in Sec.\ref{sec:FRinHarmTrap}, 
the energy dependence of the scattering length fully determines the energy spectrum.

\begin{figure}[ht]
 \centering
 \includegraphics[width=0.8\linewidth]{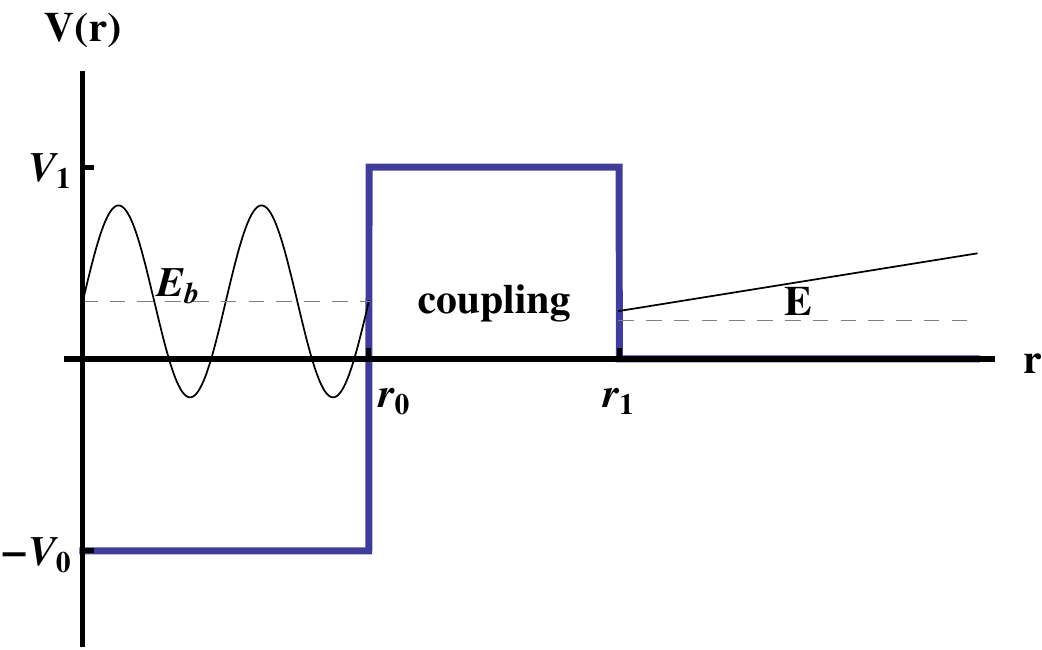}
 \caption{(Color online). Sketch of the square well potential (thick blue). The resonant character
 of the potential is due to the coupling of a bound state with energy $E_b$ (gray dashed) 
 to an unbound state with energy $E$ (gray dashed).  
 The according wave functions are sketched by black thin lines.
 For $E_b \approx E$ the scattering length 
 changes resonantly.}
 \label{fig:SquareWellPot}
\end{figure}

The square-well potential is defined as
\begin{eqnarray}
\label{eq:SqWellPot}
V(r) = \begin{cases}
-V_0 & \text{for } r\leq r_0\\
+V_1 & \text{for } r_0 < r \leq  r_1\\
0    & \text{elsewhere}\\
\end{cases}
\end{eqnarray}
with $V_0,V_1>0$ (see Fig.~\ref{fig:SquareWellPot}). 
This potential has also been used to study effects of the energy-dependence of the
scattering length on the BEC-BCS crossover \cite{cold:jens06}.
 For sufficiently large $V_0$ the potential supports a bound state behind a potential barrier 
of height $V_1$ and width $r_1-r_0$.
An atom pair that collides with an energy $E=\hbar^2k^2/(2\mu)$
scatters resonantly, if $E$ is close to the bound-state energy.

Introducing dimensionless variables $\rho = r/r_1$, $d=r_0/r_1$, $\kappa = k r_1$, 
$v_{0} = V_{0}/E_0$, and $v_{1} = V_{1}/E_0$ with $E_0 = \frac{\hbar^2}{2\mu r_1^2}$
the solution of the Schr\"odinger equation for $E>0$ is given as 
\begin{equation}
 \phi(\rho) = 
   \begin{cases}
      C \sin(k_0 \rho) & \text{ for } \rho < d \\
      A e^{k_1 \rho} + B e^{ - k_1 \rho} & \text{ for } d \leq \rho < 1 \\
      \sin\left( \kappa \rho + \varphi(k) \right) & \text{ elsewhere, } 
   \end{cases}
\end{equation}
with $k_0 = \sqrt{v_0 + \kappa^2}$ and $k_1 = \sqrt{v_1 - \kappa^2}$.

In the case of pure $s$-wave scattering one has $\kappa \ll 1$ so that one can
make, e.g., the replacements $\sin(\kappa) \rightarrow \kappa$ and $\cos(\kappa) \rightarrow 1$.
Eliminating  $A,B,$ and $C$  by demanding that the wavefunction is continuous and differentiable
the scattering length can be obtained as
\begin{equation}
\label{eq:a_sqw}
\frac{a(\kappa^2)}{r_1} \equiv -\frac{\tan \varphi}{\kappa} =
 \frac{1 +  \epsilon_{\rm res}}
      {\epsilon_{\rm res} - \kappa^2}
\end{equation}
with
\begin{align}
  \epsilon_{\rm res} &= k_1 \frac{\alpha + \beta}{\alpha -\beta}, \\
  \alpha &= e^{2 d k_1} \left[k_0 \cos (d k_0)-k_1 \sin (d k_0)\right], \\
  \beta &= e^{2 k_1}   \left[k_0 \cos (d k_0)+k_1 \sin (d k_0)\right].
\end{align}

From the functional behavior of Eq.~\eqref{eq:a_sqw} one can determine 
the corresponding parameters of the FR, i.e.\ $E_{\rm res}$, $\Delta E$, and
$\abg$.
The resonance positions of $a(\kappa^2)$ are given by the roots of 
$\kappa^2 = \epsilon_{\rm res}(\kappa^2)$.
The smallest root shall be called $\kappa_{\rm res}^2  = 
\epsilon(\kappa_{\rm res}^2)$. Hence, the resonance position evaluates to 
\begin{equation}
\label{eq:TCEres}
E_{\rm res} = E_0 \kappa_{\rm res}^2.
\end{equation}
According to Eq.~\eqref{eq:a_E_trap} the scattering length is zero if $E=E_{\rm res}-\Delta E$.
Be $\kappa_0$ the solution of $1 +  \epsilon_{\rm res}(\kappa_0) = 0$ that is closest to 
$\kappa_{\rm res}$ then 
\begin{equation}
\label{eq:TCDeltaE}
\Delta E = E_0(\kappa_{\rm res}^2 - \kappa_0^2).
\end{equation}

In order to determine the value of the background scattering length $\abg$, 
$\epsilon_{\rm res}$ is expanded linearly in $\kappa^2$ around the resonance position, yielding
\begin{eqnarray}
\label{eq:epsilon_lin}
\epsilon_{\rm res}(\kappa^2) &\approx& \kappa_{\rm res}^2 + \delta(\kappa^2-\kappa_{\rm res}^2) \\
\label{eq:delta}
\text{with}\quad \delta &=& \left.\frac{\partial \epsilon_{\rm res}}{\partial (\kappa^2)}\right|_{\kappa=\kappa_{\rm res}}.
\end{eqnarray}

\begin{figure}[ht]
 \centering
 \includegraphics[width=0.9\linewidth]{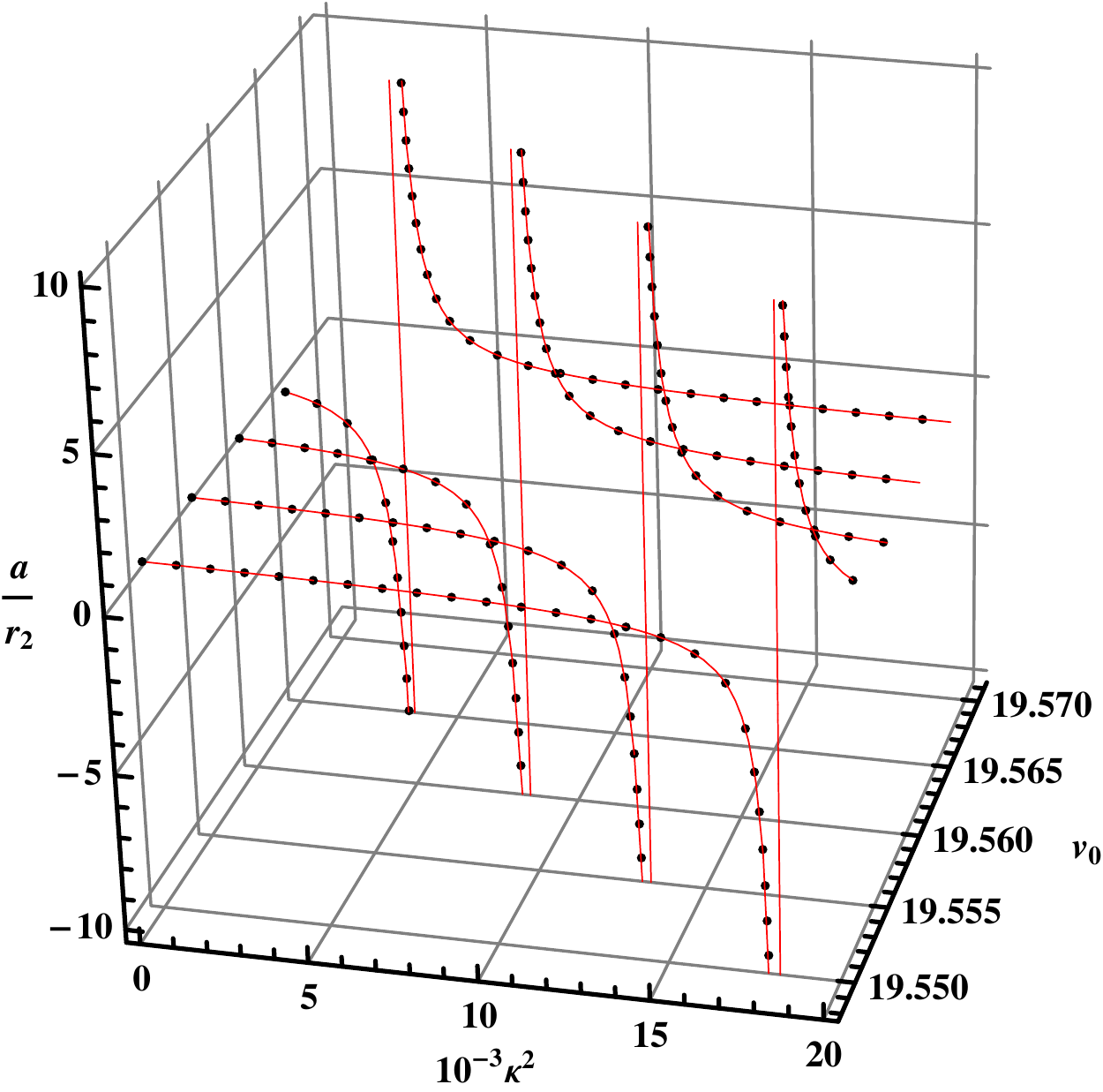}
 \caption{(Color online) Energy-dependent scattering length of the square-well potential (dots) and approximation according to Feshbach theory (thin red) for $v_1=70$ and $r_0/r_1=0.6$.}
 \label{fig:SquareWellModel}
\end{figure}

For $\kappa \rightarrow \kappa_{\rm res}$ the scattering length evaluates according to Eq.~\eqref{eq:a_sqw}
and Eq.~\eqref{eq:epsilon_lin} to
\begin{equation}
\label{eq:a_sqw_lin}
\begin{split}
\frac{a(\kappa^2)}{r_1} &=
       \frac
                   {\frac{1}{\delta-1}\left(\kappa_{\rm res}^2 +1\right)}
                   {\kappa_{\rm res}^2 - \kappa^2} 
\end{split}
\end{equation}
By comparing with the behavior of Eq.~\eqref{eq:a_E_trap} for $E\rightarrow E_{\rm res}$, 
$a = \abg \Delta E/(E_{\rm res}-E)$, one finds
\begin{eqnarray}
\label{eq:TCabg}
\abg = r_1 \frac{E_{\rm res}+E_0}{\Delta E (\delta -1)}\,.
\end{eqnarray}
For non-resonant background scattering the wavefunction simply falls off exponentially for $r<r_1$.
Therefore $a_{\rm bg}\lesssim r_1$. 
Since the potential mimics an $s$-wave resonance, the choice for $r_1$ is limited 
to $k r_1 \ll 1$ and for energies $E\approx \hbar\omega$ to $r_1 \ll a_{\rm ho}$, allowing only for rather small
positive background scattering lengths. 
On the other hand, one can freely choose $E_{\rm res}$ and $\Delta E$ by an appropriate choice of 
the parameters $v_0$ and $v_1$, respectively.
In order to also control the background scattering length one could add another
square well with $V<0$ in front of the potential in Eq.\eqref{eq:SqWellPot}. 
However, here the focus lies on the coupling to the RBS and not on the value of $\abg$.

In Fig.~\ref{fig:SquareWellModel} $a(\kappa^2)$ is shown for an exemplary square-well potential with
$d = 0.6$ and $v_1 = 70$.
The values of $a(\kappa^2)$ according to Eq.~(\ref{eq:a_sqw}) and its approximation 
\begin{equation}
a = \abg\left(1-\frac{\Delta E}{E_{\rm res}-E}\right)
\end{equation}
with the parameters according to the equations \eqref{eq:TCEres}, \eqref{eq:TCDeltaE}, and
\eqref{eq:TCabg} agree almost perfectly, showing that the square-well potential reproduces very
well the behavior of a FR. 

\section{Comparison of Bose-Hubbard model to non-perturbative calculations}
\label{sec:Comparison}

\subsection{Energy spectrum}

\begin{figure}[ht]
 \centering
 \includegraphics[width=0.9\linewidth]{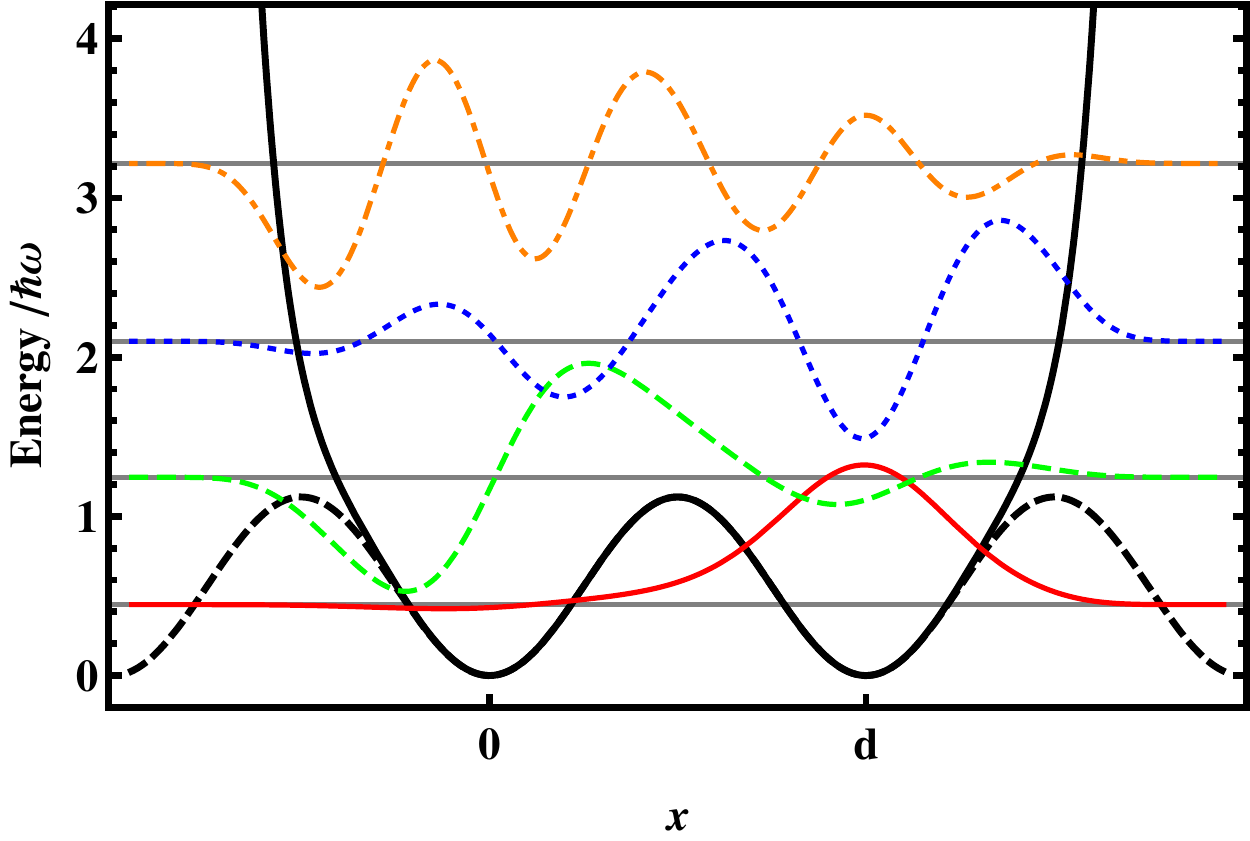}
 \caption{(Color online) Double-well potential (thick, solid) used in the {\it ab initio} calculations 
 and corresponding full lattice potential $V_{\rm L}\sin^2(k_0 x)$ (thick, dashed). 
 The Wannier functions of the atoms in the BH model are depicted for bands one to four (red solid,
 green dashed, blue dotted and orange dot-dashed) alternately for the right and the left well.
 Already above the first band they clearly probe regions where the double-well potential significantly
 differs from the full lattice potential. Horizontal lines mark the onsite energies of bands one to four.
 }
 \label{fig:WannierDoubleWell}
\end{figure}

Equipped with the possibility to model FRs with a single-channel potential
we can apply the {\it ab-initio} approach introduced in \cite{cold:gris11} to determine the
energy spectrum of two atoms at an FR in a small OL with a lattice spacing
of $d=500\;$nm.
Within the numerical approach one can expand the OL potential in all directions to some
arbitrary order. Again, to avoid unnecessary complexity the OL is expanded to harmonic order 
around $y=z=0$ in $y$ and $z$ direction and to 12-th order around $x=\pi/2$ in $x$ direction.
The lattice depth in $y$ and $z$ direction is chosen sufficiently large 
($\omega_y = \omega_z = 3.8 \omega$ where $\omega$ is the trap frequency of the harmonic 
approximation of the lattice wells in $x$ direction) 
such that excitations in these directions can be ignored. The resulting double-well potential in 
$x$ direction is shown in Fig.~\ref{fig:WannierDoubleWell}.
\begin{figure*}[ht]
 \centering
 \includegraphics[width=\linewidth]{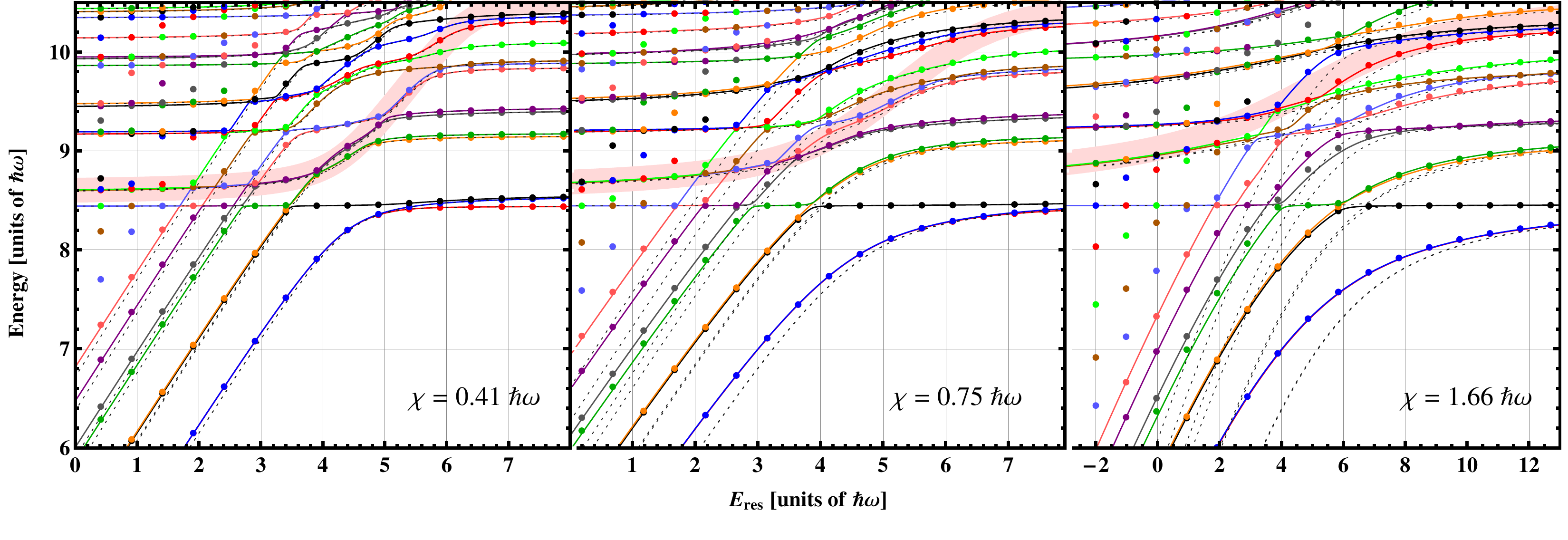} 
 \caption{(Color online) Spectra of the {\it ab-initio} calculations (dots) and the BH model with
 usage of the dressed bound-state energies and coupling (lines). Also shown are the energies of
 the undressed BH model (dotted lines). 
 The {\it ab-initio} calculations include the representation of bound states with many \cm excitations.
 Not all of these bound states are present in the BH model that only includes four Bloch bands.
 For example, in the right graph all {\it ab-initio} energies for $7.4\hbar\omega<E<8.4\hbar\omega$ 
 and $E_{\rm res}<0$ are not covered by the BH model. From left to right the parameters 
 $\abg =(88, 87, 85)\,\text{a.u.} = (9.3, 9.2, 9.0)\cdot 10^{-3}d$, 
 $\Delta E=(1.4, 4.9, 24.2)\hbar\omega$ are chosen. This corresponds to a coupling
 strength of $\chi = (0.41, 0.75, 1.66)\,\hbar\omega = (0.36, 0.67, 1.48)\,V_{\rm L}$.
 The red shading marks the energy of the repulsively interacting atoms within a single-band approximation.
 From left to right the energy of this state is significantly influenced by the bound state in the second, 
 third and fourth Bloch band demonstrating that for stronger coupling bound states in more Bloch bands have 
 to be included to obtain accurate eigenenergies.}
 \label{fig:SpectraLattice}
\end{figure*}
For large lattice depths the spectrum converges to the one of two uncoupled harmonic traps.
In order to probe the accuracy of the BH model a relatively small lattice depth of 
$V_{\rm L} = 5 E_r =1.1 \hbar\omega$
is chosen in $x$ direction.
For this low lattice depth excited states in higher Bloch bands probe parts
of the potential that significantly deviate from an ordinary lattice potential $V_{\rm L}\sin^2(k_0 x)$.
Therefore, the correct single-atom states deviate significantly from ordinary Wannier functions.
This insufficiency can be corrected for by replacing the ordinary Wannier basis by a basis constructed
from single-atom eigenstates in the double well.
For each band $p$ the left and right Wannier functions are constructed by superpositions
of the $n$-th symmetric eigenstate with energy $E_p^{(\rm even)}$ and the $n$-th anti-symmetric eigenstate
with energy $E_n^{(\rm odd)}$. The corresponding atomic Wannier functions of the first four Bloch bands are shown in 
Fig.~\ref{fig:WannierDoubleWell}. As one can see 
they are neither symmetric nor anti-symmetric so that any selection rule for the BH parameters
(such as that of the coupling between the open and the closed channel) of the OL does not apply.
The onsite energies are given as $\epsilon_n = \frac12(E_n^{(\rm odd)} + E_n^{(\rm even)})$
and the hopping parameters as $J_n=\frac12(E_n^{(\rm odd)} - E_n^{(\rm even)})$.
Furthermore, to be sure that all errors are solely due to deficiencies of the representation of the Feshbach 
resonance in the BH model also next-neighbor (background) interaction is included.

In Fig.~\ref{fig:SpectraLattice} the spectrum of the {\it ab-initio} calculation for three different 
coupling strengths is compared to the corresponding dressed and non-dressed BH spectrum. 
In contrast to Fig.~\ref{fig:spectrumBroadLattice} the trap states do not appear in energy bands
due to the reduced size of the system.
The bound states appear as duplets with one symmetric and one antisymmetric \cm excitation in $x$ direction.
Again, excited bound states in higher Bloch bands are able to couple to the
first trap state (lowest horizontal line) by next-neighbor coupling, i.e.\ the bound state couples to a state
of one atom in the same well and one in the neighboring well. For symmetry reasons only the lower bound state of 
each dublet can couple to the lowest symmetric trap state \cite{cold:stec11}.

\begin{figure}[ht]
 \centering
 \includegraphics[width=0.9\linewidth]{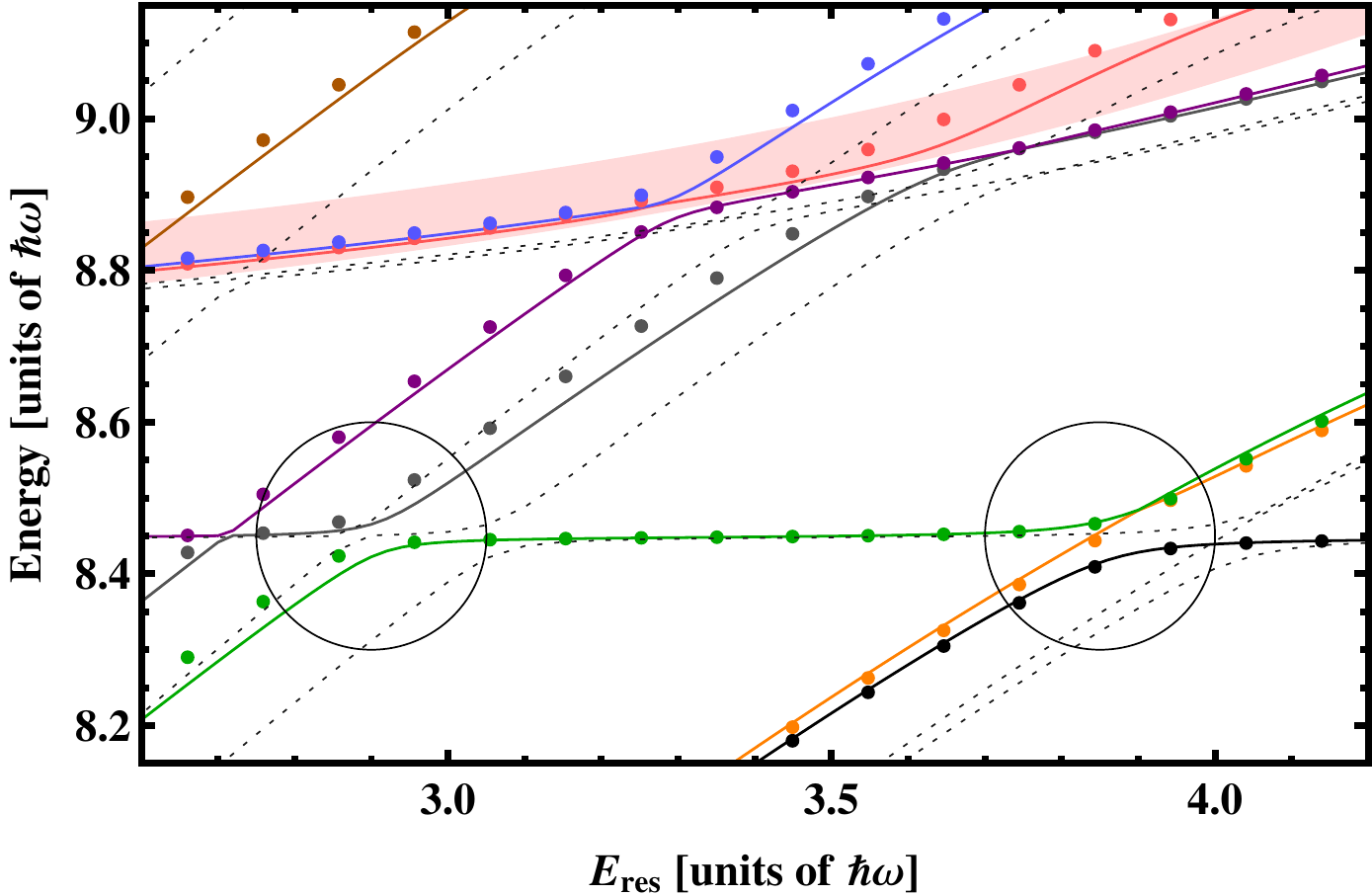}
 \caption{(Color online) Zoom on the resonance of the bound state in the second Bloch band (right circle)
 and third Bloch band (left circle) with the state of two separated atoms in the ground state for 
 $\chi = 0.75 \hbar\omega$. The splitting energies of the left resonance ($0.04\hbar\omega$) and that of
 the right resonance ($0.06\hbar\omega$) are well reproduced by the dressed BH models.}
 \label{fig:SpectrumLatticeDetail}
\end{figure}

Fig.~\ref{fig:SpectrumLatticeDetail} shows a detailed view onto two of these avoided crossings around 
$E=8.4\hbar\omega$ for a resonance energy of $E_{\rm res}=2.9\hbar\omega$ and $E_{\rm res}=3.9\hbar\omega$.
Clearly, the splitting of the avoided crossing and hence also the next-neighbor coupling strength is well 
reproduced by the dressed BH model.

Given the large degree of anharmonicity of the lattice potential
the agreement between the {\it ab-initio} spectra and BH spectra in Figs.~\ref{fig:SpectraLattice} 
and~\ref{fig:SpectrumLatticeDetail} is very good. 
The dressed bound-state energies are obtained from a harmonic approximation of the
two lattice sites. Already in the second Bloch band the potential and therefore the states and energies 
differ significantly from their harmonic counterparts (see Fig.~\ref{fig:WannierDoubleWell}).
Nevertheless, the dressed bound-state energies and the dressed coupling strength lead to
a drastic improvement of the undressed results in all three cases shown in Fig.~\ref{fig:SpectraLattice}.
In general, the dressed parameters should lead to an improvement as long as the couplings of the bound states
to trap states that probe anharmonic parts of the potential, i.e. with energies above $E=V_{\rm L}$, is negligible. 
Approximately, for $\chi \geq V_{\rm L}$ this is not the case any more since at 
the avoided crossing of the lowest bound state with the lowest trap state an energy regime above 
$V_{\rm L}$ is entered. 
Indeed, considering the spectrum with the largest coupling energy $\chi = 1.48 V_{\rm L}=1.66\hbar\omega$, 
the lowest bound-state energy of the BH model is slightly lower than that of the {\it ab-initio} calculations. 
But still the disagreement is surprisingly small.
As one can expect the correction of the bound-state energies in the third and fourth Bloch band 
is less accurate than that of the first and second Bloch band. Already for the lower coupling energies of 
$\chi = 0.36 V_{\rm L}=0.41\hbar\omega$ and $\chi = 0.67 V_{\rm L}=0.75\hbar\omega$ small disagreements 
between the corresponding eigenenergies of the {\it ab-initio} calculations and the corrected BH model appear.
 
The coupling of the two atoms in the lowest Bloch band to the bound state in the lowest Bloch band
leads to the appearance of both attractively and repulsively interacting states.
The energy of the repulsively interacting state is marked by the red shading in
Figs.~\ref{fig:SpectraLattice} and ~\ref{fig:SpectrumLatticeDetail}. 
As one can see for larger and larger coupling energy $\chi$
this state is strongly influenced by bound states in higher and higher Bloch bands.
If this energy range shall be correctly reproduced this sets a lower limit for the number of
Bloch bands that must be included in the BH model.
In Fig.~\ref{fig:SpectrumLatticeDetail} one can see that the dressed BH model reproduces correctly the
energy of the repulsively interacting state while the undressed model underestimates its energy.

As discussed above, the dressed BH model reproduces accurately the correct eigenenergies up to 
coupling energies $\chi \sim V_{\rm L}$. This corresponds usually to small up to medium FRs.
As discussed in Sec.~\ref{sec:FRinHarmTrap} a FR in a harmonic trap is broad if 
$\abg \Delta E \gg \aho \hbar\omega$. Since $\chi$ is a measure for the energy splitting of 
the avoided crossing of the lowest bound state with the first band, it is comparable to 
$\sqrt{\abg\Delta E/(\aho \hbar\omega)}\hbar\omega$ in the harmonic trap.
Therefore, a broad resonance requires $[\chi/(\hbar\omega)]^2\gg 1$. Since the BH model is valid for 
$\chi \sim V_{\rm L}$ it can only accurately describe broad FRs in a very deep lattice with 
$(V_{\rm L}/(\hbar\omega))^2 = V_{\rm L}/(4 E_r) \gg 1$.

However, for broad FRs all details of the interaction apart
from the value of the scattering length for $E\rightarrow 0$ are irrelevant (see Sec.~\ref{sec:FRinHarmTrap}). 
In this situation there is not required to explicitly include the bound state in the BH model. 
Instead, corrected BH models like the one introduced in \cite{cold:schn09} already provide accurate results.

\subsection{Time-dependent manipulations}

In the following it is studied how well the BH model can predict the dynamic behavior of the system
under the influence of some time-dependent perturbation 
\[
\oper V_{\rm pert}(t) = \sum_{i=1}^M v_{\rm pert}(\vec r_i\,) f(t)\,,
\]
which acts on each of $M$ identical atoms in the same way.

Normally, any external potential $v_{\rm pert}(\vec r_i)$ is approximately constant 
on the length scale of the bound state. Hence, the perturbation cannot couple the orthogonal 
closed and open channel states. The matrix elements of the perturbation of the closed channel 
evaluate to
\[
\begin{split}
& \QMa{\psi_b \tilde w_{i,n}}{\sum_{i=1}^M v_{\rm pert}(\vec r_i\,)}{\tilde w_{j,m} \psi_b} =
 \int d \vec R\; \int d \vec r\; |\psi_b(\vec r\,)|^2 \times \\
& \times \tilde w_{i,n}(\vec R) 
\left[v_{\rm pert}\left(\vec R + \frac12 \vec r\right) + v_{\rm pert}\left(\vec R - \frac12 \vec r\right)\right]  \tilde w_{j,m}(\vec R)\\
&\approx \int d \vec R\; \tilde w_{i,n}(\vec R) \int d \vec r\;
 |\psi_b(\vec r\,)|^2\; 2v_{\rm pert}(\vec R)  \tilde w_{j,m}(\vec R)\\
&= 2 \QMa{\tilde w_{i,n}}{\oper{v}_{\rm pert}}{\tilde w_{j,m}}
\end{split}
\]
Hence, in second quantization the perturbation is expressed as
\[
\begin{split}
\oper V_{\rm pert}(t) = f(t) \bigg(&  
  \sum_{\left<i,j\right>}\sum_{n,m=1}^N \QMa{w_{i,n}}{\oper{v}_{\rm pert}}{w_{j,m}} a_{i,n}^\dagger a_{j,m} \\
& + 2 \sum_{\left<i,j\right>}\sum_{n,m=1}^N \QMa{\tilde w_{i,n}}{\oper{v}_{\rm pert}}{\tilde w_{j,m}} b_{i,n}^\dagger b_{j,m} 
\bigg)
\end{split}
\]
As usual, only next-neighbor coupling and on-site coupling are considered and the basis is restricted to
the first $N$ Bloch bands.

\begin{figure}[ht]
 \centering
 \includegraphics[width=0.75\linewidth]{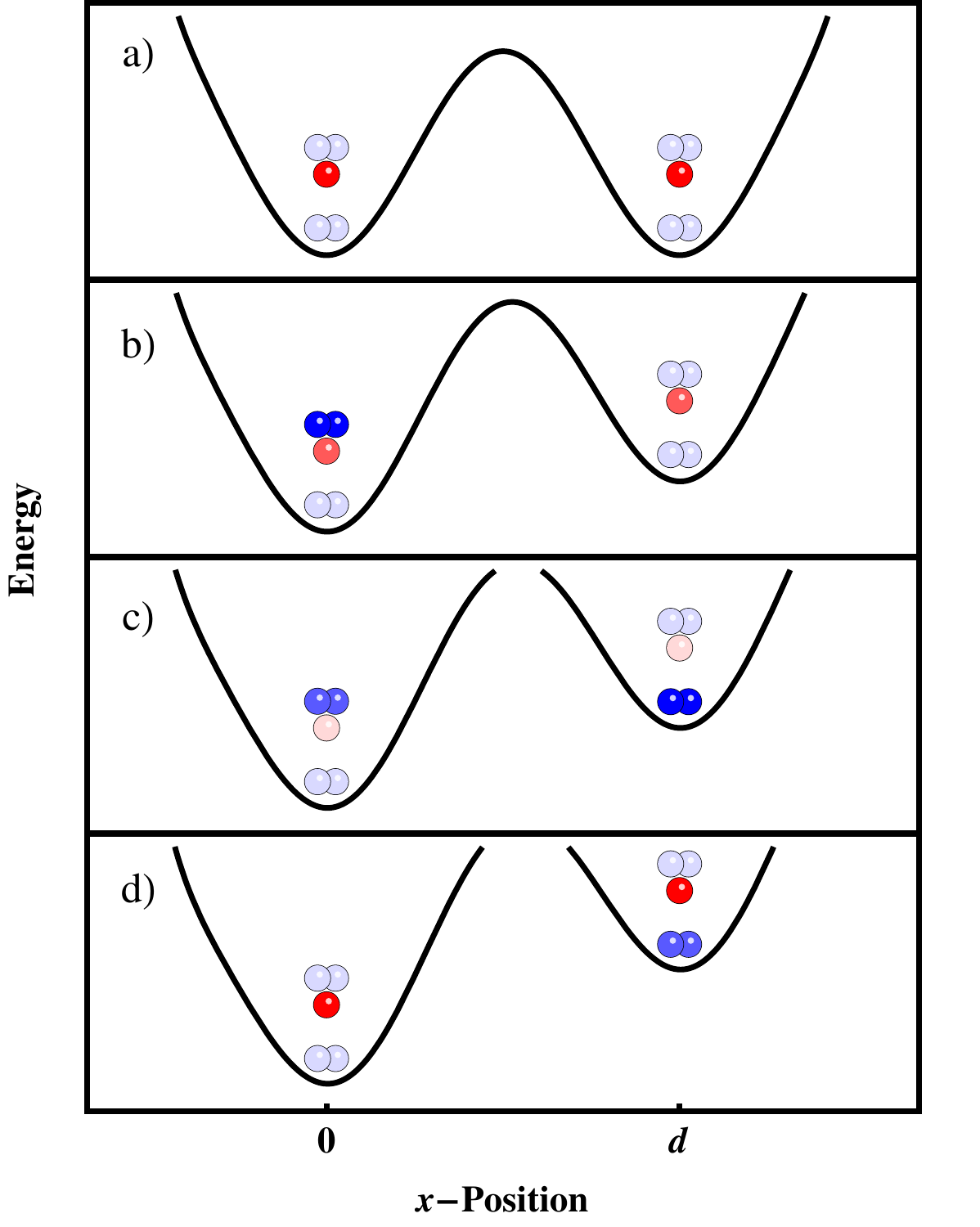}
 \caption{(Color online) Sketch of the dynamical behavior while accelerating (inclining) the double-well. 
 a) The initial state consists of separated atoms (red disks) in the ground state of the left and right well.
    The four molecular states in the \cm ground state (blue double disk below red disks) and in the first 
    excited \cm state (blue double disk above red disks) are not in resonance.
 b) Upon inclining the potential the energy of an excited molecular state in the left well (dark blue)
    comes in resonance with the energy of the separated atoms. 
    The molecular state is occupied and the \cm of the system moves to the left.
 c) After a further inclination the energy of the excited molecule in the left well comes into resonance with 
    the ground-state molecule in the right well. By occupying this state the \cm of the system moves to the right.
 d) Finally, the molecule on the right well comes into resonance with the initial state of two separated atoms and the
    \cm of the system moves again to the left.
}
 \label{fig:SketchTimeDep}
\end{figure}

In the following, the case of a linear perturbation with increasing strength, 
\[
\oper V_{\rm pert}(t) = \sum_{i=1}^N \hat x_i \lambda t   \quad ,
\]
is considered, which corresponds to an increasing acceleration of the lattice \cite{cold:schn12a}.
The dynamical behavior due to $\oper V_{\rm pert}$ is governed mainly by two effects: (i) 
The linear perturbation leads to a coupling between Wannier functions of odd and even symmetry, i.e.\ between 
bands with odd and even quantum numbers. (ii) The energy of the states at each lattice site are shifted 
proportionally to the product $\lambda d j$ and thus depend on the site number $j$.

Of course, the dynamical behavior also strongly depends on the value of the resonance energy $E_{\rm res}$. 
For the dynamical studies a resonance energy is chosen such that an inclination leads to the resonant next-neighbor
coupling of two separated atoms in the ground state to a bound state in the first and second Bloch band. 
The corresponding dynamical behavior is sketched in Fig.~\ref{fig:SketchTimeDep}.
As one can see the \cm movement of the system upon accelerating the lattice depends crucially on the
resonance energy, i.e. the energy of the RBS. Depending on the bound state and its \cm excitation that 
comes into resonance the system can move against the direction or in direction of the acceleration.
A precise representation of the system is thus necessary to predict the mobility behavior
of two atoms at a Feshbach resonance.

\begin{figure*}[ht]
 \centering
 \includegraphics[width=0.32\linewidth]{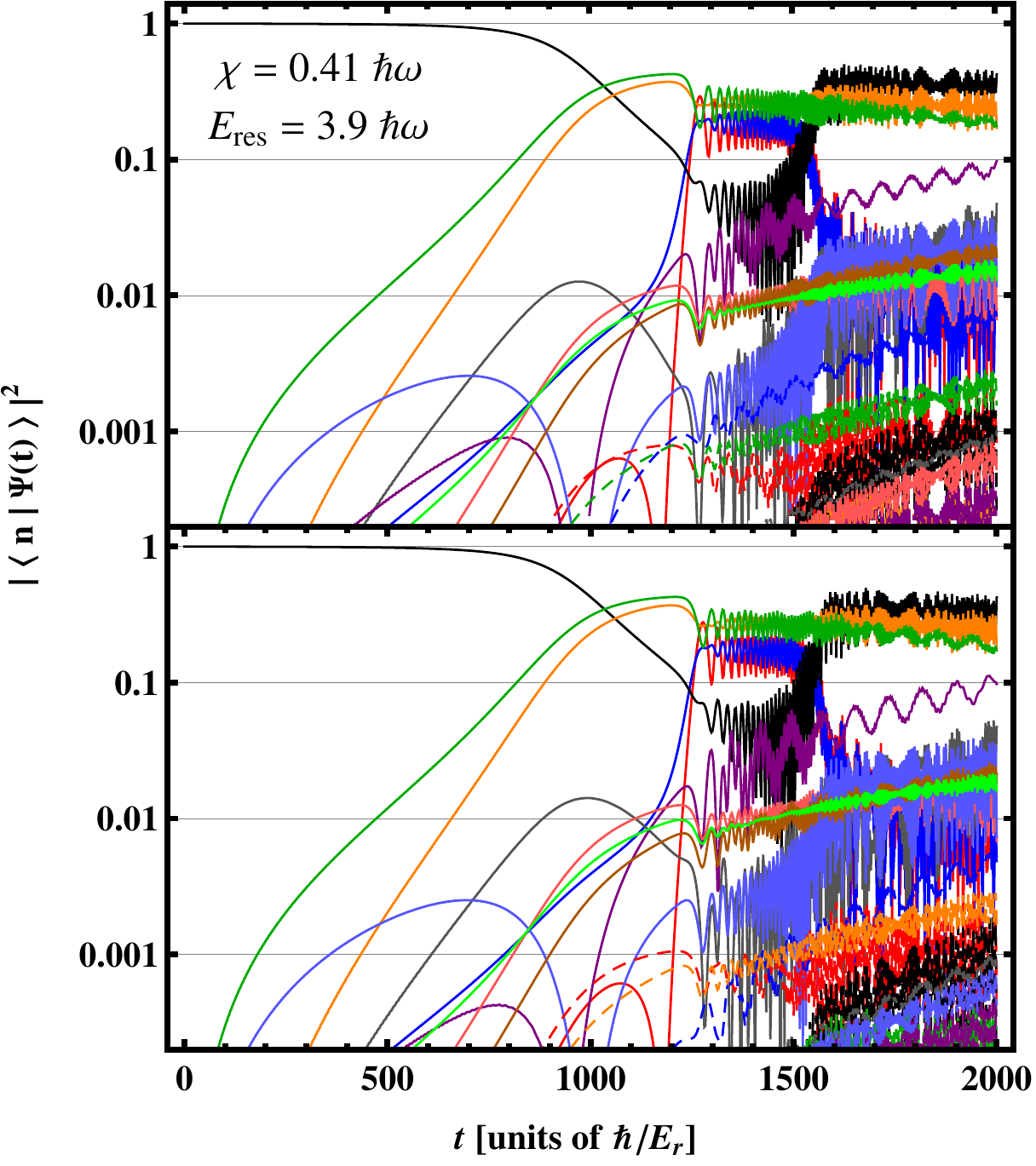}\hfill
 \includegraphics[width=0.32\linewidth]{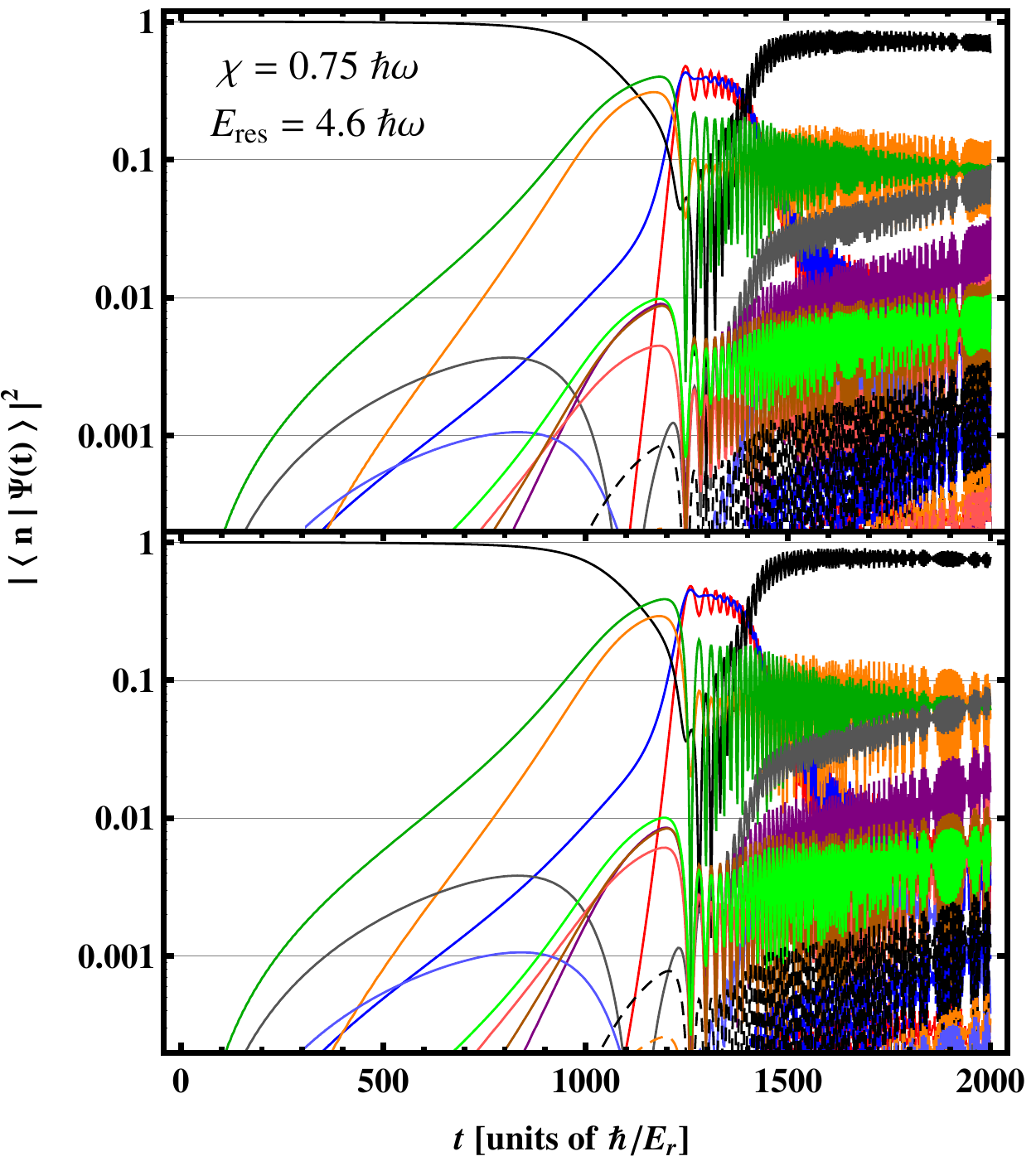}\hfill
 \includegraphics[width=0.32\linewidth]{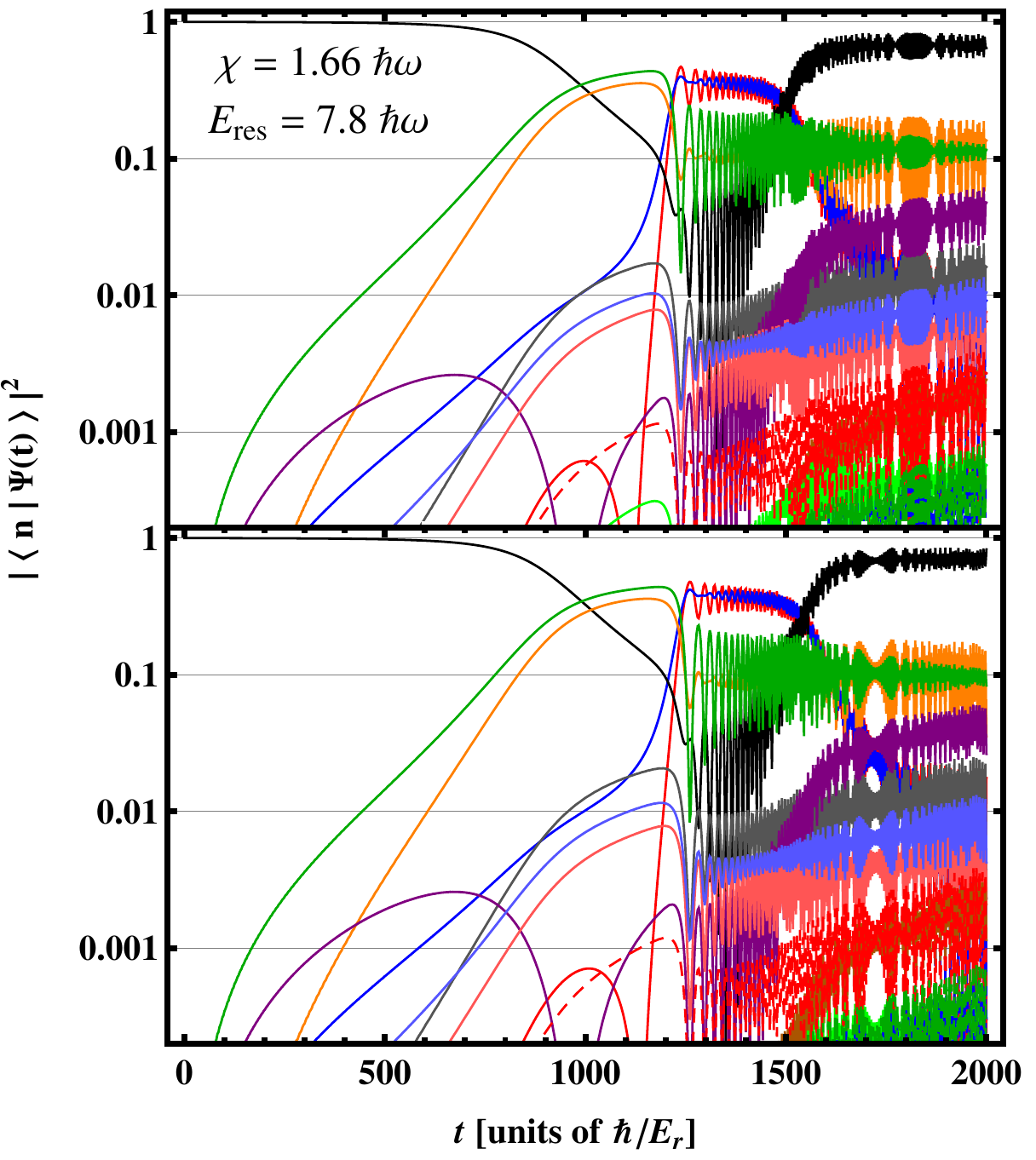}\\ 
 \hspace*{0.2cm}
 \includegraphics[width=0.3\linewidth]{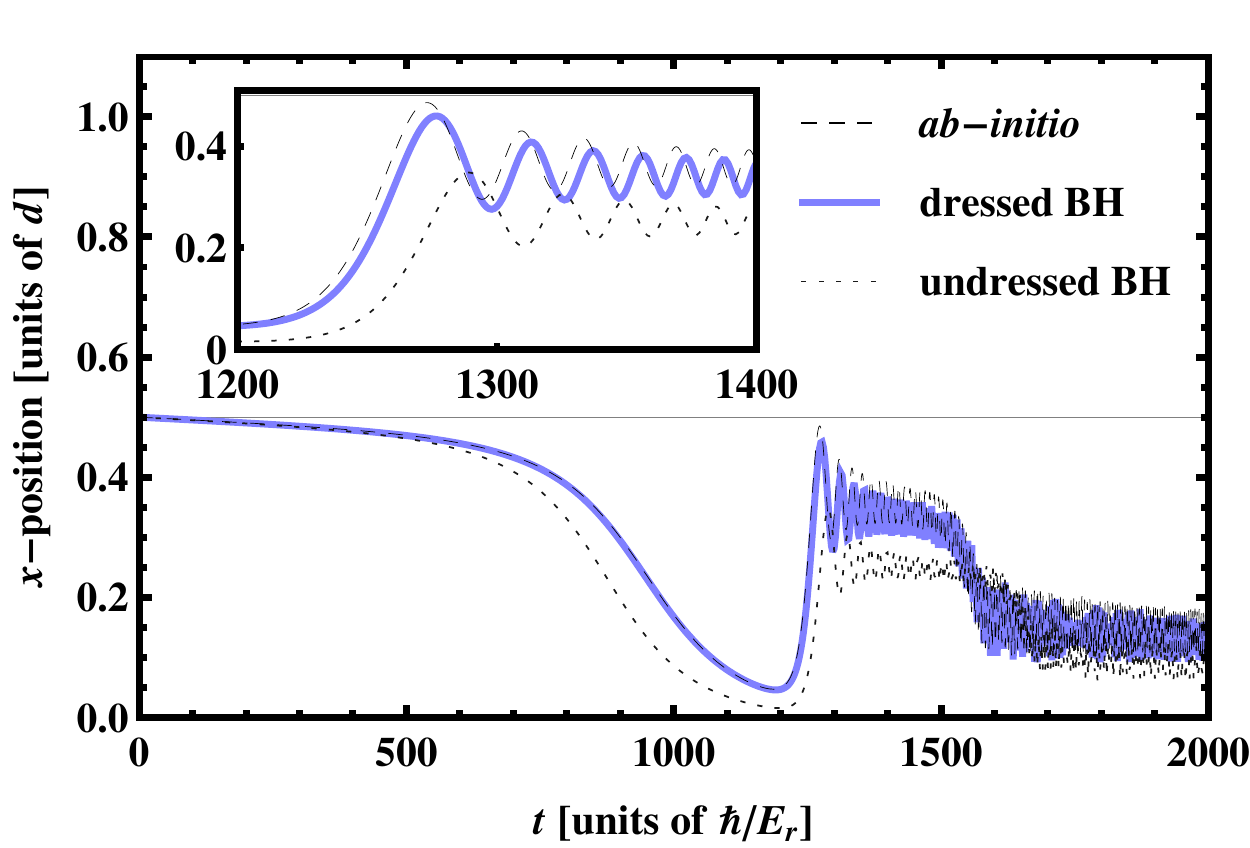}\hfill
 \includegraphics[width=0.3\linewidth]{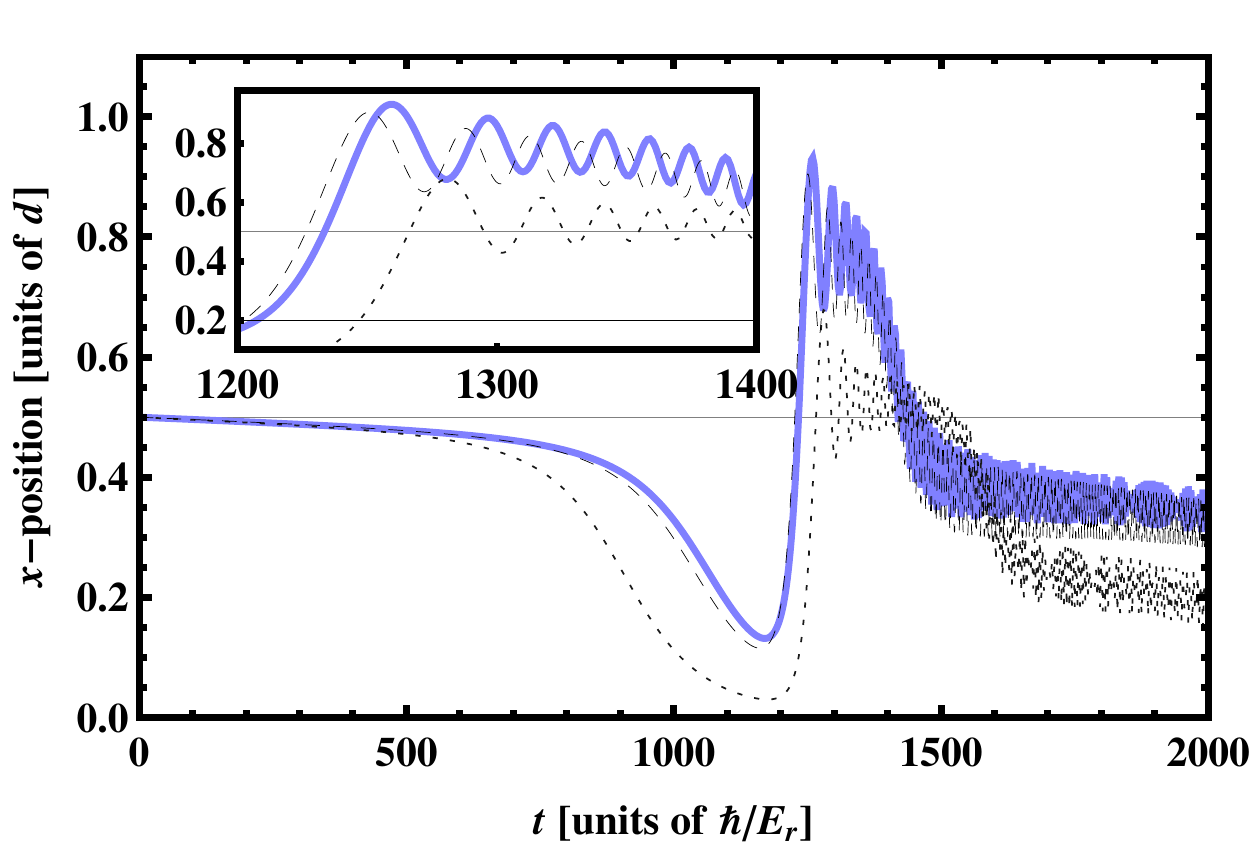}\hfill
 \includegraphics[width=0.3\linewidth]{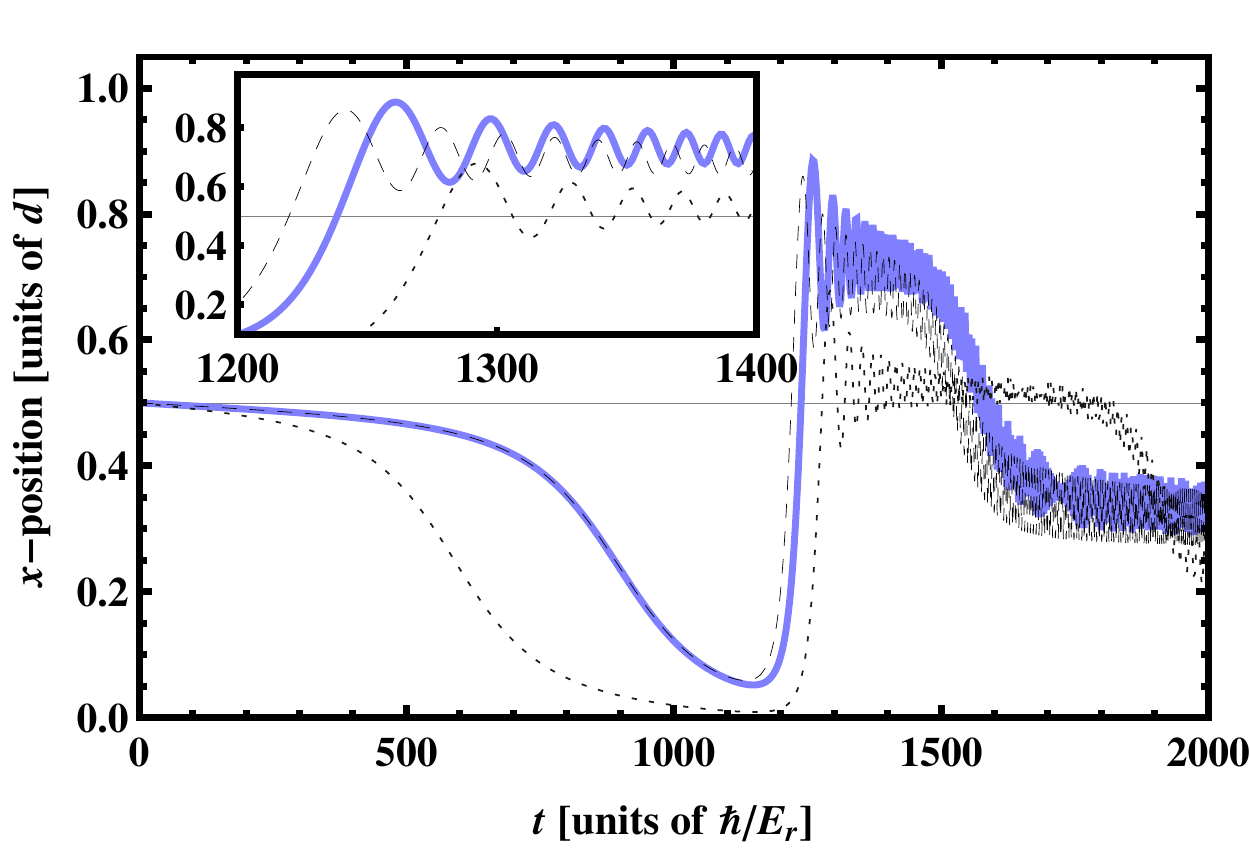}\\
 \caption{(Color online) Dynamic behavior of two separated atoms in the ground state of the double-well potential 
 during an inclination of the lattice for different coupling energies $\chi$ and resonance energies $E_{\rm res}$.
 At $t=t_{\rm end}=2000\hbar/E_r$ each atom experiences a perturbation of $\oper W=0.7\hbar \omega \hat x/d$
 which suffice to bring both the first and second bound state into resonance (see Fig.~\ref{fig:SketchTimeDep}).
 The projection of the time-dependent wave function $\ket{\Psi(t)}$ onto the 
 eigenstates $\ket{n}$ of the unperturbed system is shown in the first row ({\it ab-initio} results) and the 
 second row (results of the dressed BH model) using the same color coding as in Figs.~\ref{fig:SpectraLattice}
 and \ref{fig:SpectrumLatticeDetail}.
 In the lowest row the mean \cm position $\QMa{\Psi(t)}{\hat{X}}{\Psi(t)}$ is shown for the 
 {\it ab initio} results and the dressed and undressed BH model. The insets show a magnified region of the 
 beginning of the fast oscillations between $1200 \hbar/E_r$ and $1400 \hbar/E_r$.
 }
 \label{fig:TD_Wf_MeanX}
\end{figure*}

Fig.~\ref{fig:TD_Wf_MeanX} shows the projections $|\braket{n}{\Psi(t)}|^2$ of the time-dependent wavefunctions 
$\ket{\Psi(t)}$ onto the eigenstates $\ket{n}$ of the unperturbed system for a slow inclination with 
$\lambda = 0.0003 \frac{E_r}{\hbar} \frac{\hbar\omega}{d}$.
If the perturbation would be suddenly switched off, the projections give the probability of finding the system 
in the corresponding eigenstate.
For the same three coupling energies as shown in Fig.~\ref{fig:SpectraLattice} the qualitative agreement between
the result of the {\it ab-initio} approach (upper row) and the dressed BH model (middle row) is very good.
As is visible in Fig.~\ref{fig:SketchTimeDep}, initially the bound state in the second Bloch band is slowly occupied.
After $t \approx 1300 \hbar/E_r$ this bound state gets into resonance with the bound state in the first Bloch band 
which is then occupied. After $t \approx 1500 \hbar/E_r$ the main occupation moves back to the initial state.
Additionally to the behavior described in Fig.~\ref{fig:SketchTimeDep} the inclination leads to a strong coupling 
of the bound states in the first and second Bloch bands on each lattice site.
Due to the large energy separation of these states this coupling leads to fast oscillations of the population of the 
eigenstates.

In order to examine the quantitative agreement between the {\it ab-initio} and dressed BH results the time-dependent
\cm motion of the system $\QMa{\Psi(t)}{\hat X}{\Psi(t)}$ has been determined. As one can see in the
lower row in Fig.~\ref{fig:SketchTimeDep} the quantitative agreement is very good for the smallest coupling energy 
$\chi = 0.41\hbar\omega$. For the larger coupling energies especially the fast oscillations appearing after 
$t\approx 1200 \hbar/E_r$ are less accurately reproduced by the dressed BH model. 
The phase shift and altered frequency of the oscillations is mainly due to a small underestimation
by about 1\% of the coupling strength between the stationary eigenstates within the dressed BH model.
In contrast to the dressed BH model, the undressed BH model leads even for small coupling energies to a dynamical 
behavior significantly disagreeing from the one of the {\it ab-initio} calculations.

\section{Conclusion}
 \label{sec:Conclusion}

We have introduced a Bose-Hubbard model with dressed bound states and a dressed coupling strength, which can be used 
to accurately determine the
stationary and dynamical wavefunctions of two atoms in an optical lattice at a Feshbach resonance.
The dressed parameters, which can be straightforwardly obtained from the analytically known solution of a Feshbach 
resonance in a harmonic trap, allow one to obtain an accurate solution with including only a small number of Bloch bands.
The dressing avoids the problem that the eigenenergies, obtained by a finite expansion of the solution in single-atom 
basis states, do not converge to the correct eigenenergies in the presence of a delta-like coupling to the bound state.
Hence, the introduced method permits to determine accurate solutions without 
a regularization of the potential and a numerically demanding expansion of the solution, e.g., in Bloch functions 
or Wannier functions of many Bloch bands.
The solution of this problem should be relevant to many approaches that seek to describe strongly interacting atoms 
via a multi-band Hubbard model.

Comparisons to eigenenergies and time-dependent wavefunctions obtained from a non-perturbative approach have shown that
the method is accurate as long as the coupling energy is smaller or comparable to the lattice depth.
Furthermore, we have described a possibility to realistically mimic FRs within non-perturbative 
single-channel approaches by using a square-well interaction potential.

We believe that the approach is applicable not only to optical lattices but to various kinds of anharmonic
trapping potentials.
The introduced methods should be therefore a valuable tools for investigating the exciting physics of
Feshbach-interacting atoms in various potentials and to interpret corresponding experimental findings.

\begin{acknowledgments}
The authors gratefully acknowledge financial support by the {\it Deutsche Telekom Stiftung}, the {\it Fonds 
der Chemischen Industrie}, and the {\it Humboldt Center for Modern Optics (HZMO)}. This research was supported in part 
by the {\it National Science Foundation} under Grant No.\ NSF PHY11-25915.
\end{acknowledgments}


%

%
\end{document}